\documentclass[%
amsmath,amssymb,superscriptaddress
aps,
twocolumn,
prb,
]{revtex4-2}
\usepackage{amsmath}
\usepackage{graphicx}% Include figure files
\usepackage{bm}% bold math
\usepackage{xcolor}
\usepackage{booktabs}
\usepackage{framed}
\usepackage{multirow}
\usepackage{float}
\usepackage{hyperref}
\usepackage{everysel}
\usepackage{multirow}
\usepackage{float}
\usepackage{bigints}
\usepackage{afterpage}
\DeclareMathAlphabet \mathbfcal{OMS}{cmsy}{b}{n}
\begin{document}

%\preprint{APS/123-QED}

\title {TMDC-based topological nanospaser: single and double threshold behavior}% Force line breaks with \\

\author{Rupesh Ghimire}
\email{rghimire1@student.gsu.edu}
\affiliation{%
Center for Nano-Optics (CeNO) and Department of Physics and Astronomy, Georgia State University, Atlanta, Georgia 30303
}%
\author{Jhih-Sheng Wu}%
\email{b91202047@gmail.com}
\affiliation{%
Center for Nano-Optics (CeNO) and Department of Physics and Astronomy, Georgia State University, Atlanta, Georgia 30303
}%
%\affiliation{Department of Photonics, College of Electrical and Computer Engineering, National Chiao Tung University, Hsinchu 30010, Taiwan}
\author{Fatemeh Nematollahi}
\email{fnematallohi1@gsu.edu}
\affiliation{%
Center for Nano-Optics (CeNO) and Department of Physics and Astronomy, Georgia State University, Atlanta, Georgia 30303
}%
\author{Vadym Apalkov}
\email{vapalkov@gsu.edu}
\affiliation{%
Center for Nano-Optics (CeNO) and Department of Physics and Astronomy, Georgia State University, Atlanta, Georgia 30303
}%
\author{Mark I. Stockman}%
 \email{mstockman@gsu.edu}
\affiliation{%
Center for Nano-Optics (CeNO) and Department of Physics and Astronomy, Georgia State University, Atlanta, Georgia 30303
}%

\date{\today}% It is always \today, today,

%\twocolumn[
%\begin{@twocolumnfalse}
%\oldmaketitle
\begin{abstract}
We theoretically study a topological nanospaser, which consists of 
a silver nanospheroid and $\mathrm{MoS_2}$ monolayer flake of a circular shape. The 
metal nanospheroid acts as a plasmonic nanoresonator that supports two rotating modes, which are coupled to the corresponding valleys of $\mathrm{MoS_2}$. We apply external circularly polarized light that  selectively pumps only one of the valleys of $\mathrm{MoS_2}$. The generated spaser dynamics strongly depends on the size (radius) of the $\mathrm{MoS_2}$ nanoflake. For small radius, the system has only one spasing regime when only chirally-matched plasmon mode is generated, while at larger size of $\mathrm{MoS_2}$, depending on the pump intensity, there are two regimes. In one regime, only the chirally-matched plasmon mode is generated, while in the other regime both chirally-matched and chirally-mismatched modes exist. Different regimes of spaser operation have also opposite handedness of the far-field radiated of the spaser system. Such topological nanospaser has potential applications in different areas of infrared spectroscopy, sensing, probing, and biomedical treatment. 
\end{abstract}
%\end{@twocolumnfalse}
%]
%%%%%%%%%%%%%%%%%%%%%%%%%%%%%%%%%%%%%%%%
%%%%%%%%%%%%%%%%%%%%%%%%%%%%%%%%%%%%%%%%%%%%%%%%%
\maketitle

\section{Introduction}

The advent of spaser (surface plasmon amplificaiton by stimulated emission of radiation) stems from  a work published in 2003 \cite{Bergman_Stockman:2003_PRL_spaser}, an idea, which initiated a distinct new approach for generation and amplification of nanolocalized fields. Spaser itself has seen tremendous development due to its boundless possibilities. Such metal resonator system allows one to scale down the spatial size of a laser and, at the same time, to satisfy all the ingredients of a  perfect lasing system, i.e., optical confinement, feedback, and thermal management.  After the first experimental observation of a spaser \cite{Noginov_et_al_Nature_2009_Spaser_Observation, Oulton_Sorger_Zentgraf_Ma_Gladden_Dai_Bartal_Zhang_Nature_2009_Nanolaser}, many theoritical developments have set in to the scene \cite{ Li_Li_Stockman_Bergman_PRB_71_115409_2005_Nanolens_Spaser, Stockman_JOPT_2010_Spaser_Nanoamplifier, Fedyanin_Opt_Lett_2012_Elecrically_Pumped_Spaser, Berman_et_al_OL_2013_Magneto_Optical_Spaser, Zheludev_et_al_Nat_Phot_2008_Lasing_Spaser}, which ushered to plethora of new designs  and applications. The operational frequencies of spasers cover a wide range of  spectrum from infrared to near-untraviolet and this is one of the key factors of its rapid emergence in the optical research \cite{Zhang_et_al_Nature_Materials_2010_Spaser,  Long_et_al_Opt_Expr_2011_Spaser_1.5micron_InGaAs,  Hill_et_al_Opt_Expr_2011_DFB_SPP_Spaser, van_Exter_et_al_PRL_2013_Holy_Array_Spasing, Lu-2014-All-Color_Plasmonic, Xiong_et_al_ncomms5953_2014_Room_Temperature_Ultraviolet_Spaser, Lin_et_al_srep19887_2016_Single_Crystalline_Al_ZnO_UV_Spasers, Gwo_et_al_acsphotonics_7b00184_2017_Low_Threshold_Spasers, Song_et_al_acsphotonics_b01018_2017_Perovskite_Grating_Spaser}.

Spaser, apart from fundamental research, has proved itself viable for different potential applications as an element of  opto-electrionic systems  \cite{Stockman_2018_Patent_Processor_Communications,Thresholdless,Ning,Shane_thermal,Shen_metacavity,Ding_metallic,Ding_modulation,Dolores}, as a sensor of chemical and biological agents \cite{Cheng,Wang_high-yield,wang_lasing,zhu}, and as a biological probe \cite{ Galanzha_Nat_Comm_Spaser_biological_probe_2017,biolaser} for therapeutics and diagnostics of different diseases, e.g., cancer.

The first experimentally realized spaser is a nanoshell spaser\cite{Noginov_et_al_Nature_2009_Spaser_Observation}. It consists of a metal nanocore sorrounded by a dielectric gain medium. The metal core supports the plasmonic modes, which optically interact with the gain medium. The gain medium usually consists of dye molecules \cite{Noginov_et_al_Nature_2009_Spaser_Observation, Galanzha_Nat_Comm_Spaser_biological_probe_2017}, however, different materials such as perovskites and 2D systems can be used as a gain element\cite{Sajith_morphology, weyl_fatemeh_prb}. The size of such spaser can be as small as tens of nanometers. 

The other type of spaser that consists of a nanorod (gain medium) placed near a plasmonic metal has been proposed in Ref. \cite{Oulton_Sorger_Zentgraf_Ma_Gladden_Dai_Bartal_Zhang_Nature_2009_Nanolaser}. Such spaser is generally a micrometer in size and can be used as a good source of the far-field radiation. The operation of such plasmonic nanolasers is based on the same principles as the ones of the nanoshell spaser but with one key difference. Namely, their plasmonic modes are not the localized modes as in the case  of the nanoshell spaser but the propagating modes, which are called the surface plasmon polaritions (SSPs) \cite{Zhang_et_al_Nature_Materials_2010_Spaser, Zhang_et_al_Nano_Lett_2012_Muliticolor_Spaser, Zhang_et_al_Nat_Nano_2014_Spaser_Explosives_Detection, Xiong_et_al_ncomms5953_2014_Room_Temperature_Ultraviolet_Spaser, Ma_acsphotonics.7b00438_2017_High_Stability_Spasers_for_Sensing, Wu_et_al-2018-Advanced_Optical_Materials_2018}.

A spaser design, which is based on a periodic array of nano-resonators that have coherent extended collective modes, has been proposed in Ref.\cite{Zheludev_et_al_Nat_Phot_2008_Lasing_Spaser, Plum_Fedotov_Kuo_Tsai_Zheludev_Opt_Expr_2009_Toward_Lasing_Spaser, Zheludev_et_al_srep01237_2013_Torroidal_Lasing_Spaser}. Such spaser  consists of either an array of nanocavities embedded in a metal film  \cite{van_Exter_et_al_PRL_2013_Holy_Array_Spasing} or a periodic set of metal nanoparticles suspended in a gain medium \cite{Odom_et_al_Nature_Nano_2013_Spasing_in_Strongly_Coupled_Nanoprticle_Array}.
For a special arrangement of metal nanoparticles, the system has topologically nontrivial collective plasmonic excitations. The corresponding spaser, which is called a topological spaser of type I, 
%(in type II topological spaser the gain medium has nontrivial 
%topological properties) 
has been proposed in Ref. \cite{wu2019topological}. In such spaser, two types of metal nanoshells form the honeycomb crystal structure with the broken inversion symmetry. The collective plasmonic excitations of such system have two types of topologically nontrivial chiral modes with opposite chiralities, i.e., opposite topological charges. The modes are located at the $K$ and $K^\prime $ valleys in the corresponding reciprocal space\cite{wu2019topological}. In such topological spaser,  the nontrivial topology is introduced into the plasmonic excitations of the system.

There is also another type of topological spaser, called the topological spaser of type II\cite{topological_nanospaser_rupesh}, in which the plasmonic system is topologically trivial but the gain medium has nontrivial topology. 
The example of such gain medium is a nanopatch of a transition-metal dichalcogenide (TMDC) 
monolayer. In the reciprocal space, TMDC monolayer 
has two valleys, $K$ and $K^\prime $, which have chiral electron states and are characterized by nonzero and opposite topological charges. Plasmonic modes in such nanospaser are generated in a metal nanoparticle that is placed atop of TMDC nanopatch. Thus the topology in the system is introduced only through the gain medium. We have shown in Ref. 
\cite{topological_nanospaser_rupesh} that the plasmonic modes of such nanospaser are topologically protected and determined by the topological charge of the excited valley of TMDC nanopatch. The  topological spaser of type II has been studied in Ref. \cite{topological_nanospaser_rupesh} for small size of TMDC nanopatch, which is comparable to the size of the metal nanoparticle. 
In the present paper we extend this theoretical analysis to 
large sizes of TMDC monolayer and show that the topological nanospaser 
under this condition has very rich dynamics with different types of plasmonic modes that are generated. 

\begin{figure}[h]
\begin{center}
\includegraphics[width=.95\columnwidth]{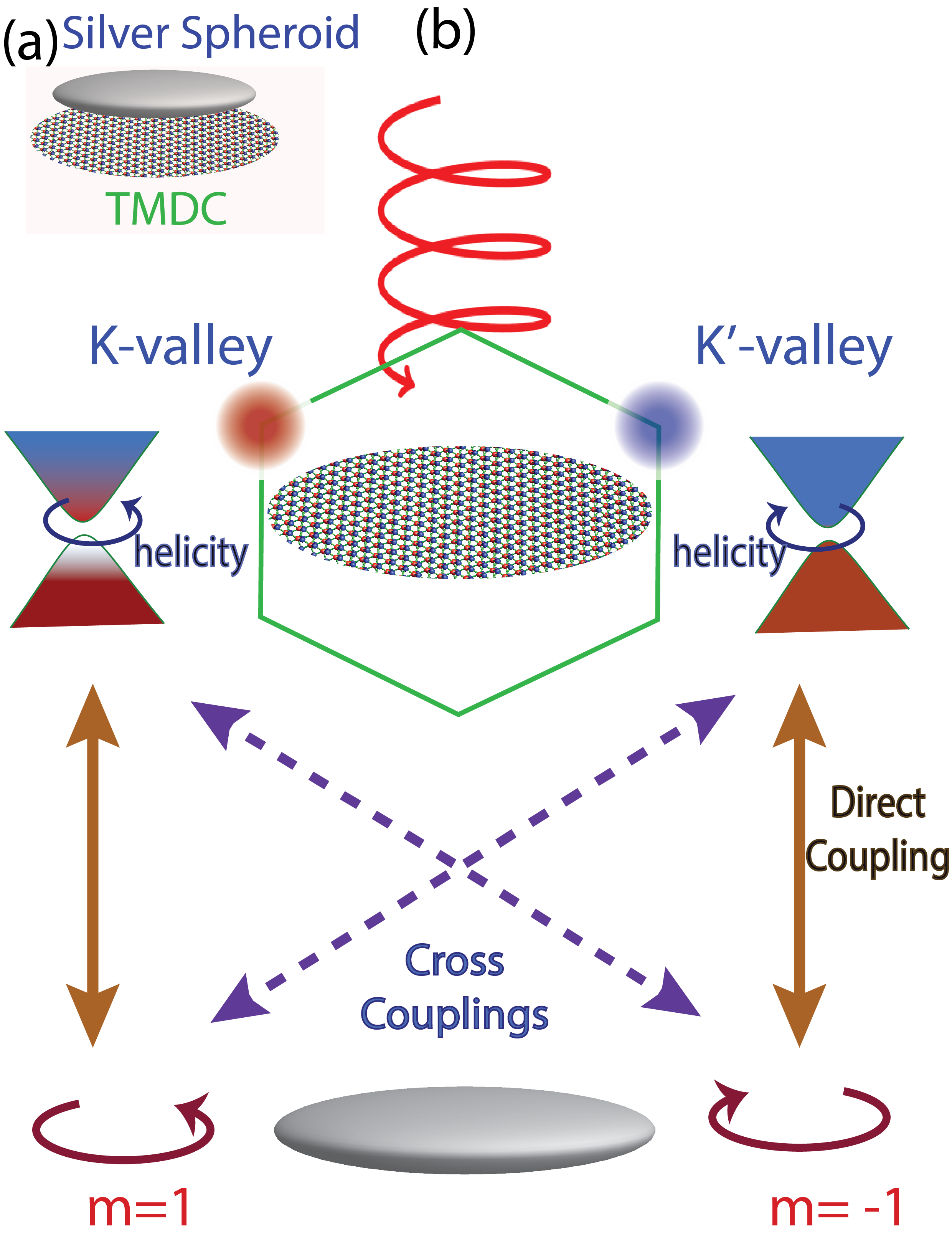}
\caption{Schematics of the topological spaser.
(a) Spaser consists of a silver nanospheroid placed on the top of TMDC nanoflake of a circular shape. The silver nanospheroid has oblate shape with radius 12 nm and height 1.2 nm. (b) Schematic of spaser operation. A circular-polarized light excites the valley with the chirality that match  the light helicity. The metal nanospheroid supports two plasmon modes with azimuthal quantum numbers $m=-1$ and $m=1$. 
The stimulated CB$\to$VB transitions at the corresponding $\mathrm{K}$ or $\mathrm{K^\prime}$ points couple to these plasmon modes through direct and cross couplings.  
}\label{fig:tmdc_spaser}
\end{center}
\end{figure}

\section{Model and Main Equations}
\label{Main_Eqs}

The topological nanospaser of type II consists of a metal nanospheroid placed on the top of a TMDC nanopatch - see Fig.\ \ref{fig:tmdc_spaser}(a). We assume that the nanospheroid has an oblate shape with the radius of 12 nm and the height of 1.2 nm. The TMDC nanopatch has a circular shape with the radius that is considered as a parameter below. The whole system is placed in dielectric medium with dielectric constant $\epsilon = 2$. 

The TMDC monolayer has two valleys, $K$ and $K^\prime $, which have opposite topological charges (chiralities). When TMDC monolayer is illuminated with circularly polarized light, only one of the valleys with the chirality that corresponds to the handedness of the circularly pilarized light is excited \cite{Cao_et_al_Nat_Commun_2012_Valley_Circular_Dichroism_MoS2, Heinz_et_al_Nat_Phys_2017_Optical_Manipulation_of_Valley_Pseudospin}. For concreteness, we assume that circularly polarized light excites only the $K$ valley [see Fig.\ \ref{fig:tmdc_spaser}(b)]. We characterize such continuous excitation by the corresponding rate $g_{\cal K}$. 

The metal nanospheroid has asimuthal symmetry and the corresponding plasmonic modes are characterized by azimuthal quantum number $m$. Because of the chiral nature of TMDC valleys, the $K$ valley  is predominantly coupled to the $m=1$ plasmon mode, while the $K^\prime $ valley is mainly coupled to the $m=-1$ mode of the nanospheroid. In Fig. 
\ref{fig:tmdc_spaser}(b), these couplings are marked as the direct couplings. The selection rules corresponding to the direct couplings are exact at the center of TMDC nanopatch ($\vec{r} =0$)  and approximate at any other points within TMDC. As a result there are also cross couplings of the plasmonic modes and TMDC valleys, which are shown schematically in 
Fig.\ \ref{fig:tmdc_spaser}(b). 

Thus in the topological nanospaser of type II, the circularly polarized light excites one of the valleys (say $K$) of TMDC monolayer (gain medium), which then excites the $m=1$ plasmonic mode through the direct coupling. Such mode is also coupled to the other valley ($K^\prime$) through the cross coupling  and excites electrons in the $K^\prime$ valley, which finally excites the $m=-1$ plasmonic mode through the direct coupling. Thus all modes of the system are coupled and the final kinetics of the system strongly depends on its parameters, e.g., the radius of TMDC nanopatch, which controls the magnitudes of both direct and cross couplings.

Within the quasistatic approximation, the plasmonic eigenmodes of nanospheroid, $\phi_m(\mathbf{r})$, satisfy the following equation\cite{Stockman:2001_PRL_Localization}
\begin{align}
\frac{\partial }{{\partial {\mathbf{r}}}}\theta \left( {\mathbf{r}} \right)\frac{\partial }{{\partial {\mathbf{r}}}}\varphi _n\left( {\mathbf{r}} \right) - s_n\frac{{\partial ^2}}{{\partial {\mathbf{r}}^2}}\varphi _n\left( {\mathbf{r}} \right) = 0 , \label{eqn_quasistatic}\\
s_n = \frac{\displaystyle{ \int_{\mathrm{All~Space}}  \theta(\mathbf{r}) |\nabla \phi_m(\mathbf{r})|^2 d^3 \mathbf{r}}}{\displaystyle{ \int_{\mathrm{All~Space}}  |\nabla \phi_m(\mathbf{r})|^2 d^3 \mathbf{r}}},
\label{s_sp}
\end{align}
where $s_n$ is the eigenvalue of the $n$th mode, $\mathbf{\theta}$ is a characterstic function which is 1 inside  and 0 outside the metal. For the oblite nanospheroid, which has azimuthal symmetry, the eigenmodes  are characterized by 
 two spheroidal quantum numbers\cite{Willatzen_Voon_2011_Book_Boundary_Problems}: multipolarity $l=1,2,\dots$ and azimuthal or magnetic quantum number $m=0,\pm1,\dots$. In our case the relevant modes are the dipole modes with $l=1$ and $m=\pm1$.

The Hamiltonian of our system has the following form 
\begin{align}
H=H_{SP}+ H_{gain}+ H_{int},\label{eqham}
\end{align}
where $H_{SP}$ is the Hamiltonian of surface plasmons (SP) (nanospheroid), $H_{gain}$ is the Hamiltonian of TMDC nanopatch (gain medium), and $H_{int}$ describes the interaction between the plasmonic system and the gain medium.

The Hamiltonian of SP, $H_{SP}$, in the quantum mechanical description, is given by the following expression
\begin{align}
H_{SP}=\hbar\omega_\mathrm{sp}\sum_{m=\pm1}  \hat{a}_m^{*}\hat{a}_m^{},
\end{align}
where $\omega_\mathrm{sp}$ is the SP frequency, $\hat{a}_m^{\dagger}$ and $\hat{a}_m^{}$ are creation and annihilation operators of the corresponding plasmon with quantum number $m=\pm 1$.  Then the electric field operator is \cite{Bergman_Stockman:2003_PRL_spaser, Stockman_JOPT_2010_Spaser_Nanoamplifier}
\begin {align}
%\mathbf{F}_m(\mathbf{r})&= - A_\mathrm{sp} \nabla \phi_m(\mathbf{r})(\hat{a}_m +\hat{ a}_m^{\dagger}),
\mathbf{F}_m(\mathbf{r},t)&=-\sum_{m=1,{-1}}  A_\mathrm{sp} (\nabla \phi_m(\mathbf{r}) \hat{a}_m  e^{-i\omega t} +\nabla \phi_m^{*} (\mathbf{r}) \hat{ a}_m^{*}  e^{i\omega t}),\label{fieldeq}\\
A_ \mathrm{sp} &=\sqrt{\frac{4\pi 
\hbar s_n}{\epsilon_d s^\prime_n}}~
\mathrm{~~and~~}
s_n^{\prime}\equiv \frac{d\mathrm{Re}[s(\omega)]}{d\omega}\Big|_{\omega=\omega_{\mathrm{sp}}}.
%\mathrm{s(\omega)}=\frac{\epsilon_{d}}{\ 
%\epsilon_{d}-\epsilon_{m(\omega)}}, \label{eq4}
\end{align}
The electric field depends on the position within 
TMDC monolayer through the position dependent plasmonic mode, $\phi_m(\mathbf{r})$. The interaction Hamiltonian describes the dipole coupling of the plasmonic excitations and the electron system in TMDC monolayer and  is given by the following expression 
\begin{align}
H_{int}=- \nu_{\mathbf{K}}\sum_{\mathbfcal{K}=\mathbf K, \mathbf K^\prime}\int d^2\mathbf{r}\sum_{m=\pm1}\mathbf{F}_m(\mathbf{r})\hat{\mathbf{d}}_\mathbfcal K(\mathbf r)~,
\end{align}
where $\mathbfcal{K}$ is the valley index, $\mathrm{K}$ or $\mathrm{K^\prime}$, $\nu_\mathbfcal{K}$ is the density of electron states at the $\mathbfcal{K}$ valley, which can be found from the experimental data\cite{Salehzadeh_et_al_ACSPublications_Optically_Pumped_2015_MoS2, Wu_et_al_Nature_2015_Monalayer}:  
$  \nu_\mathbfcal{K} =7.0\times10^{12}$~cm$^{-2} $.
Here the interband dipole matrix element, $\mathbf{d}_{\mathbfcal{K}}$, between the VB and CB states of TMDC, is proportional to the non-Abelian Berry connection
\begin{eqnarray}
\mathbf{d}_{\mathbfcal{K}}& =&e\mathbfcal A^\mathrm{(cv)}(\mathbf k)~, 
\nonumber\\
\mathbfcal A^\mathrm{(cv)}(\mathbf k)&=& i\left.\left\langle u_{\mathrm c\mathbf{k}} \left|
\frac{\partial}{\partial \mathbf{k}} \right|u_{\mathrm v\mathbf{k}}\right\rangle\right|_{\mathbf{k}=\mathbfcal{K}}~,
\end{eqnarray}
where $u_{\mathrm c \mathbf{k}}$ and $u_{\mathrm v \mathbf{k}}$ are the lattice Bloch functions.

The optical transitions  in TMDC nanopatch, which are the transitions between the valence band (VB) and the conduction band (CB), are characterized by the transiton frequency $\omega_{21}$, which is the same for both $K$ and $K^\prime$ valleys due to time reversal symmetry of the system. 
Below we choose the parameters of nanospheroid for which  the 
 SP frequency $\omega_{sp}$ is equal to the transition frequency $\omega_{21}$.

The spaser dynamics is described within the semi-classically approach. Within this approach the plasmonic system is treated classically, i.e.,  the creation and annihilation operators ($\hat{a}_m=a_m$ and $\hat{a}_m^\dag=a_m^\ast$) are considered as complex numbers, and, at the same time, the electron system in TMDC  (gain medium) is described quantum mechanically using the density matrix $\hat{\rho}_{\mathbfcal{K}}(\mathbf{r},t)$. In the rotating wave approximation (RWA) \cite{Bloch_Siegert_PhysRev.57.522_1940, Agarwal_PhysRevA.4.1778_1971} the density matrix has the following form
\begin{align}
\hat{\rho}_{\mathbfcal{K}}(\mathbf{r},t) & =\left(
\begin{array}{cc}
\rho_{\mathbfcal{K}}^\mathrm{(c)}(\mathbf{r},t) & \rho_{\mathbfcal{K}}(\mathbf{r},t) e^{i\omega t}\\
\rho_{\mathbfcal{K}}^{*}(\mathbf{r},t)  e^{-i\omega t} & \rho_{\mathbfcal{K}}^\mathrm{(v)}(\mathbf{r},t)
\end{array}
\right).
\end{align}

Then the corresponding semiclassical equations can be derived from Hamiltonian (\ref{eqham}), 
\begin{align}
\dot{a}_m=[i&(\omega - \omega_\mathrm{sp}) - \gamma_\mathrm{sp}]{a}_m +
\nonumber\\
& i \nu_\mathbfcal{K}\int_S d^2{\mathbf{r}}\sum_{\mathbfcal{K}}\rho^{*}_{\mathbfcal{K}}(\mathbf{r})\tilde{\Omega}_{m,\mathbfcal{K}}^{*}(\mathbf{r})~,
\label{12}
\end{align}
\begin{align}
\dot{n}_{\mathbfcal{K}}(\mathbf{r})=&-4\sum_{m=1,-1} \mathrm{Im} \left[{\rho}_{\mathbfcal{K}}(\mathbf{r}) \tilde{\Omega}_{m,\mathbfcal{K}}(\mathbf{r})a_{m}^{}\right] +\nonumber\\
&g_\mathbfcal{K} \left[1-{n}_{\mathbfcal{K}}(\mathbf{r})\right]-
%\nonumber\\& 
\gamma_{2\mathbfcal{K}}(\mathbf{r}) \left[1 +{n}_{\mathbfcal{K}}(\mathbf{r})\right],~\label{13}~
\end{align}
\begin{align}
\dot{\rho}_{\mathbfcal{K}}(\mathbf{r})=[-i(\omega-\omega_{21})-&\Gamma_{12}] {\rho}_{\mathbfcal{K}}(\mathbf{r})+
\nonumber\\
&i n_{\mathbfcal{K}}(\mathbf{r})\sum_{m=1,{-1}} \tilde{\Omega}_{m,\mathbfcal{K}}^{*}a_{m}^{*}~,
\label{14}
\end{align} 
where $S$ is the area of TMDC nanoflake, $\Gamma_{12}$ is the polarization relaxation rate, $g_\mathbfcal K$ is the pumping rate of the $\mathbfcal K$ valley , and $\tilde{\Omega}_{m,\mathbfcal{K}}(\mathbf{r}) $ is the Rabi frequency, 
\begin{align}
\tilde{\Omega}_{m,\mathbfcal{K}}(\mathbf{r})&= -\frac{1}{\hbar} A_\mathrm{sp} \nabla \phi_m(\mathbf{r})\mathbf{d}_\mathbfcal{K}^\ast~.
\label{Rabi}
\end{align}
Here the population inversion, $n_{\mathbfcal{K}}$, is defined as
\begin{equation}
n_{\mathbfcal{K}} \equiv \rho_{\mathbfcal{K}}^\mathrm{(c)} - \rho_{\mathbfcal{K}}^\mathrm{(v)}~,
\end{equation}
and the spontaneous emission rate, $\gamma_{2\mathbfcal{K}}(\mathbf{r})$, of the SPs is \cite{Stockman_JOPT_2010_Spaser_Nanoamplifier}
\begin{align}
\gamma_{2\mathbfcal{K}}(\mathbf{r}) = \frac{2(\gamma_\mathrm{sp}+\Gamma_{12})}{(\omega_\mathrm{sp}-\omega_{21})^2+(\gamma_\mathrm{sp}+\Gamma_{12})^2}\sum_{m=1,-1}\left| \tilde{\Omega}_{m,\mathbfcal{K}}(\mathbf{r})\right|^2
\end{align}

\begin{figure*}[htpb]
\centering
\includegraphics[scale=.8]{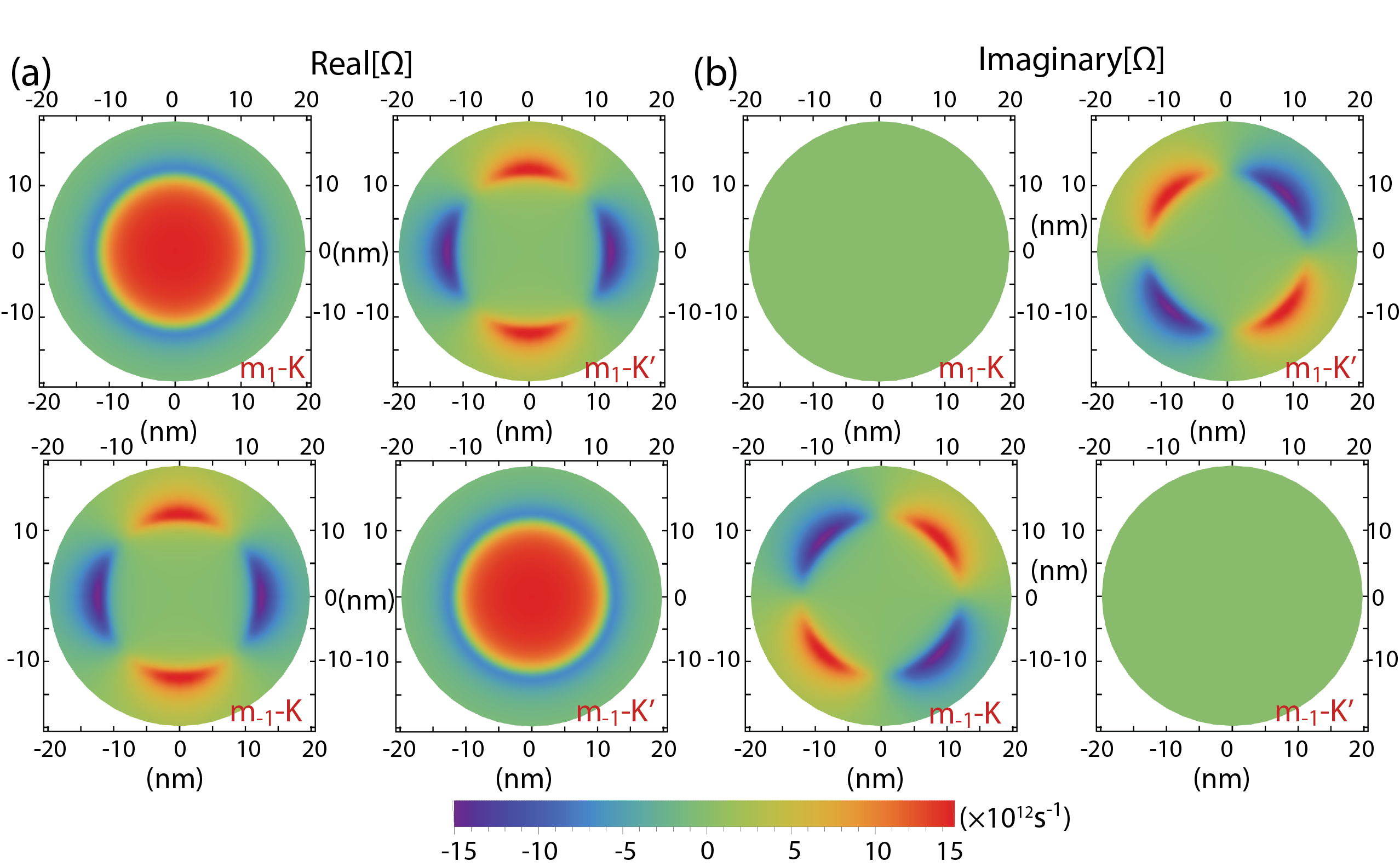}
\caption{ The real part (a) and imaginary part (b) of the Rabi frequency. The Rabi frequency  determines the coupling of the plasmon mode $m$ and the $K$ or $K^\prime $ valleys of TMDC. The radius of the metal spheroid is $a=12$ nm.}
\label{SP_Coupling}
\end{figure*}

\section{Results and Discussions}

\subsection{Parameters of topological spaser}
\label{Params}

The system consists of three components: a silver nanospheroid, a TMDC monolayer nanoflake, and a surrounding dielectric medium. A silver nanospheroid has an oblate shape with the radius of $12$ nm and the height of $1.2$ nm. These are the semi-major ($a$) and semi-minor ($c$) axes of an oblate system. We choose a $\mathrm{MoS_2}$ monolayer  as a TMDC material. The  $\mathrm{MoS_2}$ nanoflake has a circular shape and is placed on the top of the nanospheroid. Below we change the radius of $\mathrm{MoS_2}$ nanoflake. Both the nanospheroid and $\mathrm{MoS_2}$ monolayer are suspended in dielectric medium with the permittivity of $\epsilon_d=2$.

The height of nanospheroid, $c$, is chosen in a such way that the SP frequency $\omega_{sp}$ matches the dipole transition frequency $\omega_{21}= \Delta_g/\hbar$ of $\mathrm{MoS_2}$. Here $\Delta_g$ is the bandgap of $\mathrm{MoS_2}$ monolayer. 
The band structure of $\mathrm{MoS_2}$ was calculated within the three-band tight binding model\cite {Liu_et_al_PRB_2014_Three_Band_Model}. From this model we obtained both the bandgap and the dipole matrix elements between the conduction and valence bands at the $K$ and $K^\prime $ points: $\Delta_\mathrm g=1.66$ eV; 
$\mathbf d_\mathbf K=17.7 ~\mathbf e_+~\mathrm D$, and $\mathbf d_{\mathbf K^\prime}=17.7 ~\mathbf e_-~ \mathrm D$, where $\mathbf e_\pm=\left(\mathbf e_x\pm i \mathbf e_y\right)/\sqrt2$ are the chiral unit vectors. 

The other parameters, which have been used in our calculations, are the coherent relaxaton rate $ \Gamma_{12}$ = 15.1 $\mathrm{ps}^{-1}$ for $\mathrm{MoS_2}$  monolayer and the plasmon relaxation rate for silver  nanoparticle, $\gamma_\mathrm{sp}$ =13.4 $\mathrm{ps}^{-1}$. We also assume that only the $K$ valley is pumped by external circularly polarized light, i.e., $g_{\mathbf{K}}=g$ and  $g_{\mathbf{K^\prime}}=0$. With the above parameters, from the nonlinear system of Eqs. (\ref{12})-(\ref{14}), we obtain both the the stationary solution and the kinetics of topological spaser for given initial conditions.

\subsection {The dynamics of topological spaser}

We solve the system of Eqs. (\ref{12})-(\ref{14}) numerically with the given initial conditions, which are  $N_{m=1}=
|a_1|^2 = 9$ ($a_1 = 3$), i.e., there are nine $m=1$ plasmons, and the conduction band populations of both $K$ and $K^\prime$ valley are zero. We obtain the spaser dynamics for different values of the gain, $g$, and different radii of 
TMDC nanoflakes. Here we assume that only one valley, i.e., the $K$ valley, is pumped by the circularly polarized light.

The dynamics of the spaser is completely determined by the  
coupling of plasmonic excitations and the TMDC system. Such coupling is characterized by the Rabi frequency,
$\tilde{\Omega}_{m,\mathbfcal{K}}(\mathbf{r})$, which is a function of the radius vector, $\mathbf{r}$, within TMDC nanoflake and also depends on the type of the plasmon, $m=1$ or $m=-1$, and the valley of TMDC, $K$ or $K^\prime$. 
The real and imaginary parts of $\tilde{\Omega}_{m,\mathbfcal{K}}(\mathbf{r})$ are shown in Fig. \ref{SP_Coupling} for different combinations of $m$ and TMDC valley. The radius of metal nanospheroid, which is 12 nm in the $x$-$y$ plane, determines two different dependences of $\tilde{\Omega}_{m,\mathbfcal{K}}(\mathbf{r})$. 
If $r< 12$ nm, then the Rabi frequency, both the real and imaginary parts, is isotropic function of radius. It mainly follows the "angular momentum" selection rule, i.e., $m=1$ is coupled to the $K$ valley, while $m=-1$ is coupled to the $K^\prime $ valley. This selection is exact at $r=0$, but for $r>0$ it is a good approximation. 

For $r>12$ nm, i.e., a point is outside the metal nanospheroid in the $x$-$y$ plane, the plasmonic electric field has a dipole nature. 
As a result, $\tilde{\Omega}_{m,\mathbfcal{K}}(\mathbf{r})$ behaves completely differently. Namely, both $\tilde{\Omega}_{m=1,K^\prime }$ and 
$\tilde{\Omega}_{m=-1,K}$ are large, while $\tilde{\Omega}_{m=-1,K^\prime }$ and $\tilde{\Omega}_{m=1,K}$ are small. Also, because of the dipole nature of the plasmonic electric field, the Rabi frequency 
acquires strong angular dependence of type $\exp (2i\varphi)$ - see 
Fig. \ref{SP_Coupling}, where $\varphi $ is the polar in-plane angle.

Thus, from the properties of the Rabi frequency, we can conclude that 
if the radius of TMDC nanoflake is less than 12 nm, then $m=1$ plasmon mode is mainly coupled to the TMDC valley of the same chirality, i.e, the $K$ valley, while $m=-1$ plasmon mode - to the $K^\prime $ valley. But if the radius of TMDC is greater than 12 nm, then $m=1$ and $m=-1$ plasmon modes are coupled to both $K$ and $K^\prime $ valleys. The larger the radius of the TMDC flake, the stronger the coupling of the plasmonic mode to the TMDC valley of opposite chirality.

\begin{figure}[h]
\begin{center}
\includegraphics[width=.99\columnwidth]{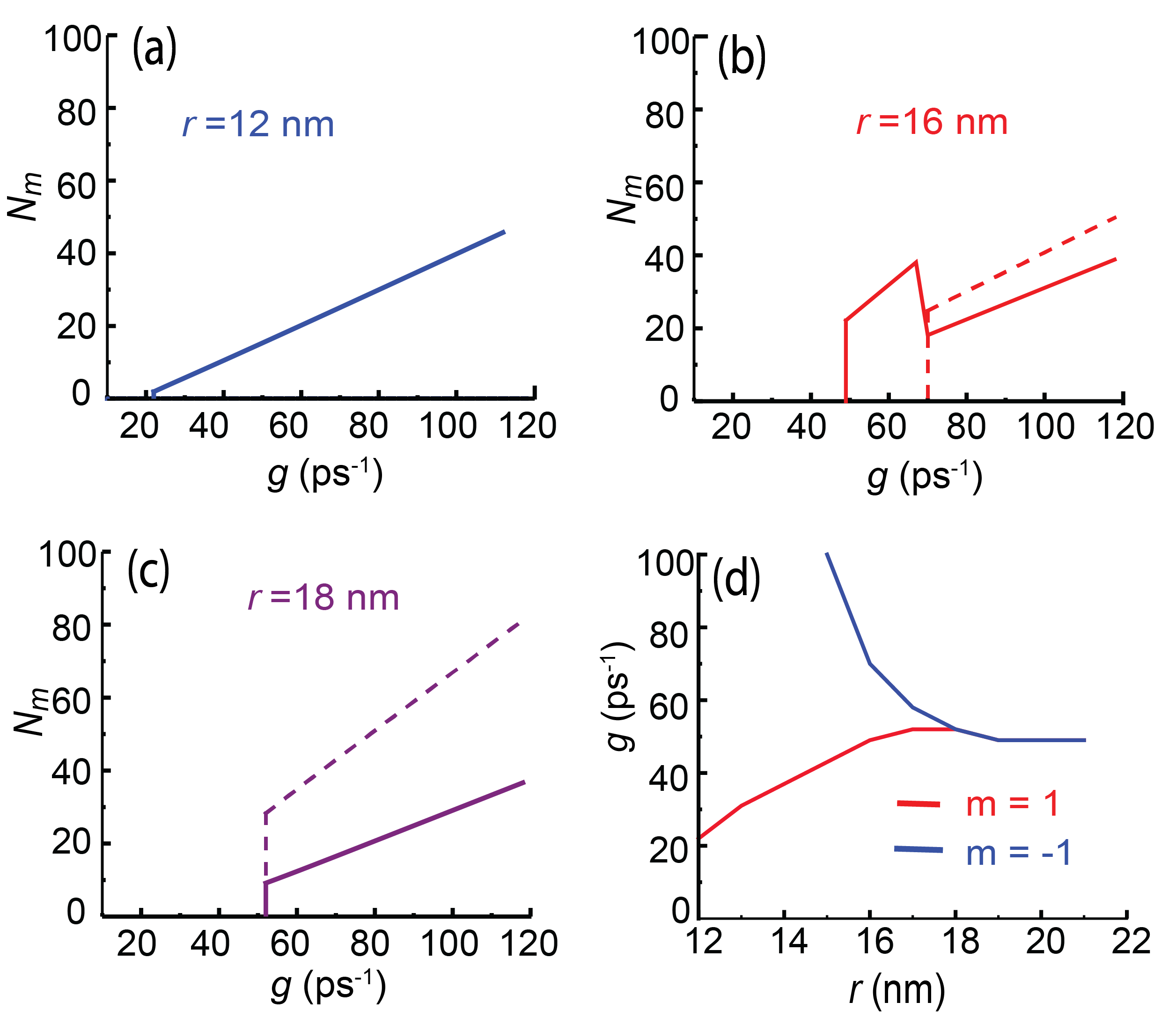}
\caption{(a)-(c) Number of plasmons, $N_m$, as a function of gain, $g$. The solid and dashed lines correspond to the plasmons with $m=1$ and $m=-1$, respectively. The radius of TMDC nanoflake is (a) 12 nm, (b) 16 nm, and (c) 18 nm. (d) The topological spaser thresholds as a function of  radius of TMDC nanoflake. If $g_{th,2}>g> g_{th,1}$ then only $m=1$ plasmon mode exists in the stationary regime, while if $g>g_{th,2}$ then both modes $m=1$ and $m=-1$ are generated. 
}\label{CW_kinetics}
\end{center}
\end{figure}

The number of plasmons, $N_m =|a_m|^2$, as a function gain, $g$, is shown in Fig.\ \ref{CW_kinetics} (a)-(c) for three different radii of TMDC nanoflake. The solid and dashed lines correspond to $m=1$ and $m=-1$ plasmons, respectively. If $r=12$ nm then only co-rotating ($m=1$) plasmon mode, i.e., the mode that is strongly coupled to the excited $K$ valley, 
is generated. There is a characteristic spaser threshold, 
$g_{th}\approx 20$ $ps^{-1}$, when the plasmon mode starts generating.  

For a larger radius, $r=16$ nm, see Fig.\ \ref{CW_kinetics}(b), the system show different behavior. Now, the $K$ valley is coupled to both co-rotating $m=1$ and counter-rotating $m=-1$ modes (although the coupling to the counter-rotating mode is still relatively weak). There are two thresholds, $g_{th,1}$ and $g_{th,2}$. At lower threshold, 
$g_{th,1}\approx 49$ $ps^{-1}$, only one mode, $m=1$, is generated, while at larger 
threshold, $g_{th,2}\approx 70$ $ps^{-1}$, both plasmon modes, $m=1$ and $m=-1$, cogenerated. At the second threshold the energy is transferred from the $m=1$ mode to $m=-1$ mode so the number of $m=1$ plasmons decreases. 
Another unique feature of the second regime, $g> g_{tr,2}$, is that there are more
counter-rotating plasmons than the co-rotating ones, $N_{-1}>N_1$. 

When the radius of the TMDC nanoflake increases even more, $r=18$ nm, two thresholds, $g_{th,1}$ and $g_{th,2}$, merge into a single one, $g_{th,1}=g_{th,2}\approx 52$ $ps^{-1}$, at which  two plasmon modes are generated simultaneously - see Fig.\ \ref{CW_kinetics}(c). Similar to a smaller radius, the number of counter-rotating plasmons is more than the number of co-rotating ones. 

The dependence of two thresholds, $g_{th,1}$ and $g_{th,2}$, on the TMDC radius, $r$, is shown in Fig.\ \ref{CW_kinetics}(d). At $r\lesssim 15 $ nm, there is only one spaser regime when only co-rotating mode is generated. At $15 nm \lesssim r \lesssim 17 nm$ there are two thresholds and the system can generate either one plasmon mode, $m=1$, or two plasmons modes, $m=1$ and $m=-1$, depending on the gain, $g$. At $r> 17$ nm, two thresholds merge into a single one and there is only one regime with two generated plasmon modes.

\begin{figure}[h]
\centering
\begin{center}
\includegraphics[width=.99\columnwidth]{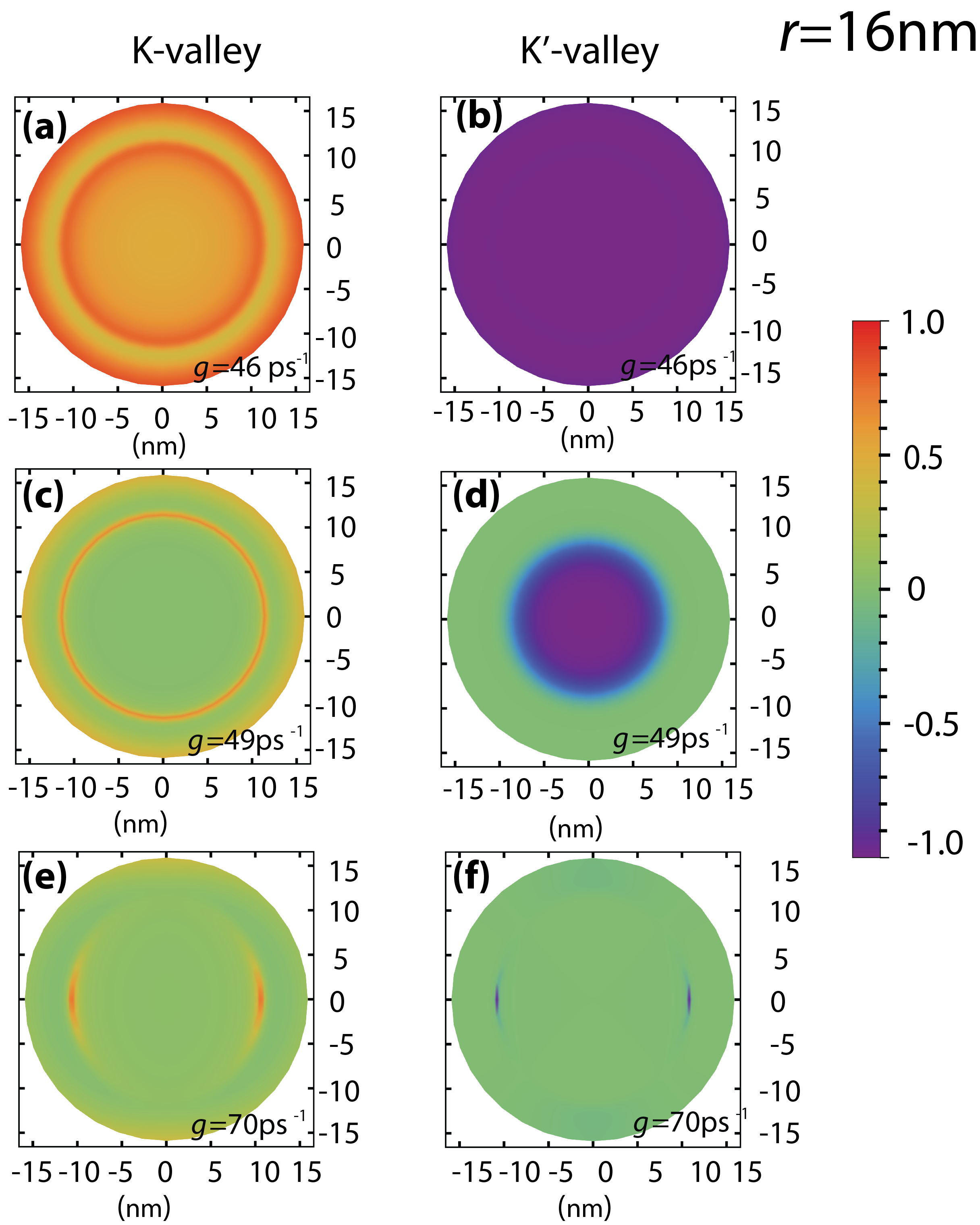}
\caption{Inversion population of $K$ and $K^\prime$ valleys of 
 $\mathrm{MoS_2}$ nanoflake with the radius of 16 nm. 
The gain is (a),(b) 46 $ps^{-1}$, (c),(d) 49 $ps^{-1}$, and 
(e),(f) 70 $ps^{-1}$. The panels (a), (c), and (e) correspond to the $K$ valley, while the panels (b), (d), and (f) describe the $K^\prime $ valley. 
}\label{inv16}
\end{center}
\end{figure}

To illustrate different regimes of spaser dynamics, we show in Fig. 
\ref{inv16} the distributions of the population inversions at the $K$ and $K^\prime $ valleys for the radius of TMDC flake of 16 nm and different values of gain. If the gain is less than the first threshold, $g< g_{th,1}$, then no plasmons are generated and the population inversion of the $K$ valley is close to one, while the population inversion of the $K^\prime $ valley is exactly -1, i.e., the valence band is completely occupied and the conduction band is empty. This case is shown in Fig.\ \ref{inv16}(a)-(b).

In Fig. \ref{inv16}(c)-(d) the gain is greater than $g_{th,1}$ but less than $g_{th,2}$. In this case only one plasmon mode, $m=1$, is generated. The distribution of population inversion is isotropic and it is close to zero for $r<12$ nm for the $K$ valley and for $r>12$ nm for the $K^\prime $ valley, which illustrates strong coupling of these spatial regions to the $m=1$ mode -see also Fig.\ \ref{SP_Coupling}.

Another possibility, when the gain is greater than $g_{th,2}$, is shown in Fig. \ref{inv16}(e)-(f). Under this condition, both $m=1$ and $m=-1$ plasmon modes are generated. They are coupled to both valleys at all spatial regions ($r<12$ nm and $r>12$ nm), as a result, the population inversions for both $K$ and $K^\prime$ valleys are close to zero. Because of coexistence of two plasmon modes, the resulting electric field show interference features and the corresponding population inversion distribution is anisotropic - see Fig. \ref{inv16}(c)-(d).

\begin{figure}
\centering
\begin{center}
\includegraphics[width=.95\columnwidth]{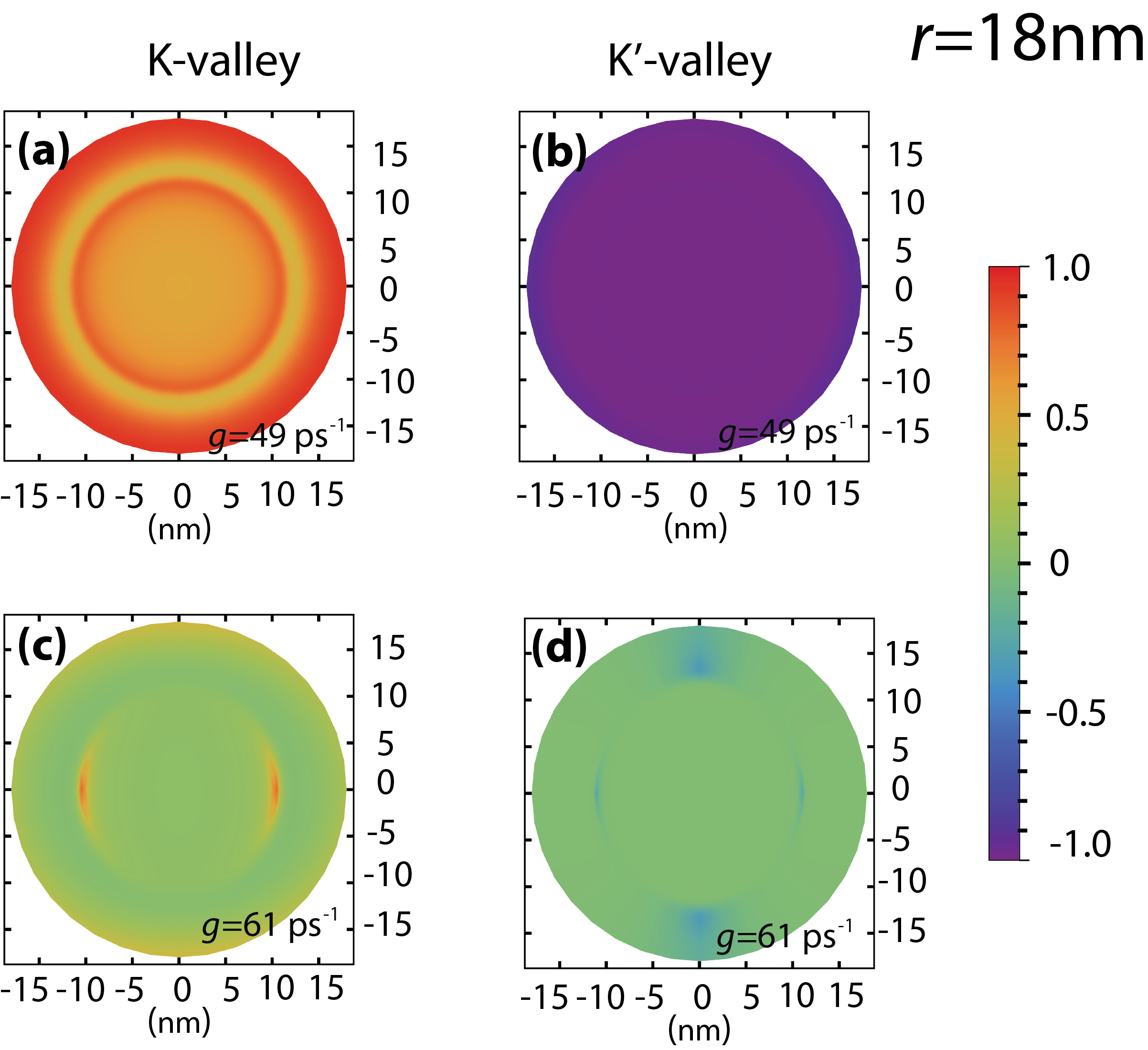}
\caption{Inversion population of $K$ and $K^\prime$ valleys of 
 $\mathrm{MoS_2}$ nanoflake with the radius of 18 nm. 
The gain is (a),(b) 49 $ps^{-1}$ and (c),(d) 61 $ps^{-1}$. The panels (a), (c) correspond to the $K$ valley, while the panels (b), (d) describe the $K^\prime $ valley. }
\label{inv18}
\end{center}
\end{figure}

The large radius of TMDC nanoflake, $r=18$ nm, is illustrated in Fig. \ref{inv18}. In this case there is only one threshold. 
If the gain is less than the threshold, see Fig. \ref{inv18}(a)-(b), then no plasmons are generated and the conduction band of the $K$ valley is highly populated, while the population inversion of the $K^\prime$ valley is -1 at all spatial points. If the gain is greater than the threshold, see 
Fig. \ref{inv18}(c)-(d), then two plasmon modes, $m=-1$ and $m=1$, are 
generated. The population inversion is close to zero for both $K$ and $K^\prime $ valleys. The population inversion distribution is also anisotropic, which is due to interference of the plasmonic fields of two modes. 

\begin{figure}
\centering
\begin{center}
\includegraphics[width=.99\columnwidth]{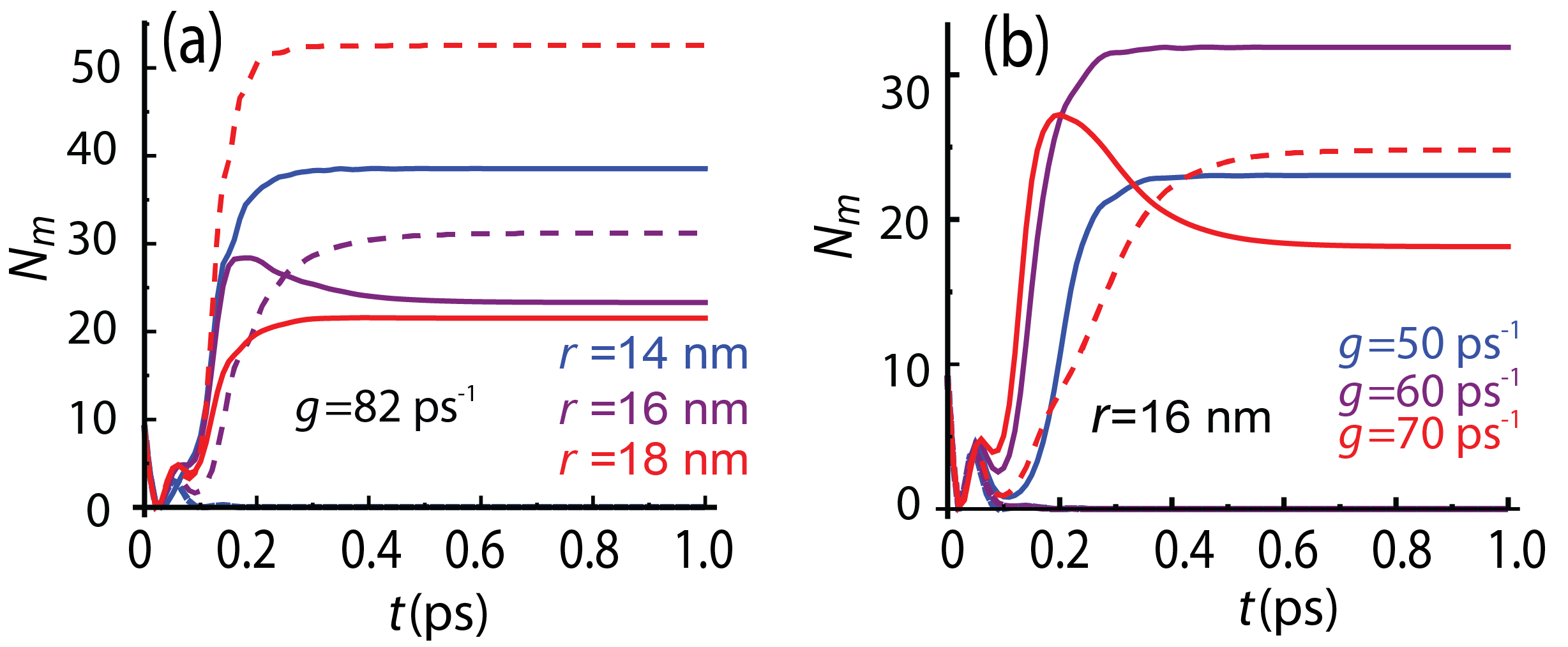}
\caption{The number of surface plasmons, $N_m$, as a function of time $t$ for topological spaser with MoS$_2$ nanoflake as a gain medium.  
The solid and dashed lines correspond to $m=1$ and $m=-1$ plasmons, respectively. The initial number of plasmons is $N_{1}=N_{-1}=9$. 
(a) The gain is $g=82$ $ps^{-1}$ and the radii of TMDC nanoflake are 14 nm, 16 nm, and 18 nm. The corresponding lines are shown by different colors as marked in the panel. (b) The radius of TMDC nanoflake is 16 nm and the gain is $50$ $ps^{-1}$, $60$ $ps^{-1}$, and $70$ $ps^{-1}$. 
The corresponding lines are shown by different colors as marked in the panel.}
\label{dynamics}
\end{center}
\end{figure}

The temporal dynamics of the topological spaser is shown in Fig.\ \ref{dynamics}. The initial number of plasmons is nine for both modes, $m=1$ and $m=-1$. In Fig.\ \ref{dynamics}(a), the gain is fixed, $g=82$ $ps^{-1}$, and the results are shown for different radii of TMDC nanoflake. 
For all parameters, the number of plasmons, $N_1$ and $N_{-1}$, show similar initial dynamics (at $t\lesssim 0.15$ ps). Namely, first, both $N_1$ and $N_{-1}$ sharply decrease to almost zero values, then show small oscillations and finally monotonically increase to the stationary values. At the stationary stage, for small radius of TMDC nanoflake, $r=14$ nm, only $m=1$ is generated, while for large radius, $r=16$ nm or $r=18$ nm, both modes, $m=1$ and $m=-1$, are generated - see Fig.\ \ref{dynamics}(a).

This property is also illustrated in Fig. \ref{dynamics}(b), in which the radius is fixed, $r=16$ nm, and the gain is varied. For $g=50$ $ps^{-1}$ and $g=60$ $ps^{-1}$, which are less then the second threshold, only $m=1$ plasmons are generated in the stationary regime. For large gain, $g=70$ $ps^{-1}$, mode $m=-1$ coexists with $m=1$ plasmonic mode. 

%The presence of the counter-rotating plasmons can be attributed to the fact that the plasmonic fied created by the co-rotating plasmons outside the footprint of the spheroid causes the polarization of $\mathrm{K^\prime}$ valley. This polarization serves as a fuel and allows a pathway for a similar process of plasmonic generation as for the co-rotating mode. These results calculated are for the initial plasmon number of 9 in each mode which we can refer to a seed value that can allow both modes to sustain. Anything lesser than a this seed population of plasmons might affect the occurance of a plasmon of a certain mode. It is to be noted that the seed value, however, doesn't change the saturation value(final value) of the plasmons generated\cite{topological_nanospaser_rupesh}. A striking observation in the calculation is merging of a pump threshold at a higher values of radii($\geq$17 nm) for two modes. This can be best explained by analyzing the the spatial inversion $ n_{\mathbfcal{K}}$ distribution within a given $valley$ $\mathcal{K}$. 

%\begin{figure}
%\centering
%\begin{center}
%\includegraphics[width=.99\columnwidth]{diffg.png}
%\caption{Schematic of CW spaser operation.
%(a) . (b) 
%}\label{fig:tmdc00_spaser}
%\end{center}
%\end{figure}

%Similar nature of changes in inversion is seen for gain with the radius of 18 nm with one contrasting difference -- the intermediate inversion chart  Fig.\ \ref{inv16}(c)-(d) is absent in Fig.\ \ref{inv18}. The larger gain radius in the later case has a relatively different effect of plasmonic field on the $valleys$ compared to the smaller gain size. Thus, the interference of the field produced isn't the same anymore as we can see from the comparision between Fig.\ \ref{inv16}(f) and Fig.\ \ref{inv18}(d) where the bluish hotspots about y-axis is absent in 16 nm gain. This pattern can be observed for gain with all other radii $\geq$17 nm. The result is a locking of both the plasmonic modes about a unique threshold as seen in Fig.\ \ref{CW_kinetics}(c)-(d).

\subsection{
Far-Field Radiation}
\label{Radiation}

 Although it is not its primary role, the topological nanolaser can be used as a miniature source of far-field radiation, which is due to the oscillating electric dipole of the spaser system.
 The induced electric dipole moment of the system is the sum of two contributions: the dipole moment of TMDC nanoflake, $\mathbf{d_{tmdc}}$, and the dipole moment of the metal nanospheroid, $\mathbf{d_{metal}}$,
\begin{equation}
\mathbf{d_{total}} = \mathbf{d_{tmdc}} + \mathbf{d_{metal}}.
\end{equation}

The dipole moment of TMDC nanoflake can be expressed in terms of the non-diagonal part of the density matrix of TMDC system 
\begin{align}
\mathbf{d_{tmdc}} &=\sum_{\mathbfcal{K}=\mathbf{K},{\mathbf{K}^\prime}}(\rho_{\mathbfcal{K}}\mathbf{d_{\mathbfcal{K}}}e^{i\omega t}+\rho_{\mathbfcal{K}}^{*}\mathbf{d_{\mathbfcal{K}}^{*}}e^{-i\omega t})\nonumber
\\&=(\rho_{\mathbf{K}}\mathbf{d_{K}}+\rho_{\mathbf{K^\prime}}\mathbf{d_{K^\prime}})e^{i\omega t}+(\rho_{\mathrm{\mathbf{K}}}^{*}\mathbf{d_{\mathbf{K}}^{*}} \nonumber \\&~~~~~~+\rho_{\mathrm{K^\prime}}^{*}\mathbf{d_{K^\prime}^{*}})e^{-i\omega t}. \label{eqtmdc}
\end{align}

The dipole moment of the metal nanospheroid can be found from the known 
electric field [see Eq. \eqref{fieldeq}] inside the metal,  
\begin{equation}
\mathbf{d_{metal}}= \int_V \frac{\mathrm{Re}[\epsilon_{metal}-\epsilon_{d}]}{4\pi}  \mathbf{F_m(\mathbf{r},t)}\,\mathbf{dr} .\label{eq5}
\end{equation}
Here the integral is calculated over the volume of the nanospheroid. The electric field, $\mathbf{F_m(\mathbf{r},t)}$, inside the metal depends on the number of $m=1$ and $m=-1$  plasmons.

The $x$ and $y$ components of the total dipole moment, which is the sum of  Eqs.\ \eqref{eqtmdc} and \eqref{eq5}, can be expressed in the following forms 
\begin{align}
\mathbf{d_{total,{x}}}&=\mathrm{B_1} e^{i\omega t}+\tilde{\mathrm{B_1} }e^{-i\omega t}, \label{19}\\
\mathbf{d_{total,{y}}}&=\mathrm{C_1} e^{i\omega t}+\tilde{\mathrm{C_1} }e^{-i\omega t}, \label{20}
\end{align}
where,
\begin{align}
\mathrm{B_1}&=-\kappa  \mathrm{A}_{\mathrm{sp}}\mathrm{ E_{0} }\mathrm {V}(\hat{ a}_1^{*}+\hat{ a}_{-1}^{*})+\mathrm{f}_{\mathbf{K}}\mathrm{d_{0}}+\mathrm{f}_{\mathbf{K^\prime}}\mathrm{d_{0}}, \label{21}\\
\mathrm{C_1}&=i\kappa \mathrm{A}_{\mathrm{sp}}\mathrm{ E_{0} }\mathrm {V}(\hat{ a}_1^{*}-\hat{ a}_{-1}^{*})+i\mathrm{f}_{\mathbf{K}}\mathrm{d_{0}}-i\mathrm{f}_{\mathbf{K^\prime}}\mathrm{d_{0}},\label{22}
\end{align}
\begin{equation}
\kappa = \frac{\mathrm{Re}[\epsilon_{metal}-\epsilon_{d}]}{4\pi} 
\end{equation}
\begin{align}
\mathrm{f}_{\mathbf{K}}=- {\nu}&
\sum_{\mathrm{S}}\frac{in_{\mathbf{K}}(\mathbf{r})\sum_{m=1,{-1}} \tilde{\Omega}_{m,\mathbf{K}}^{*}a_{m}^{*}}{-(\omega-\Delta_\mathrm g)+i\Gamma_{12}},\label{23}\\
\mathrm{f}_{\mathbf{K^{\prime}}}=- {\nu}&\sum_{\mathrm{S}}\frac{in_{\mathbf{K^{\prime}}}(\mathbf{r})\sum_{m=1,{-1}} \tilde{\Omega}_{m,\mathbf{K^{\prime}}}^{*}a_{m}^{*}}{-(\omega-\Delta_\mathrm g)+i\Gamma_{12}},\label{24}\\
 \mathbf{d_{K}}&=\mathrm{d_{0}}\mathbf{e_{+}} ~~\& ~~\mathbf{d_{K^\prime}}=\mathrm{d_{0}}\mathbf{e_{-}}.
\end{align}
The derivation of  Eqs.\ \eqref{21}- \eqref{24} is given in the Supporting Materials. The main contribution to the total dipole moment comes from the metal nanospheroid, for which the corresponding dipole moment is almost two order of magnitude larger than the dipole moment of TMDC nanoflake. 

\begin{figure}
\centering
\begin{center}
\includegraphics[width=.99\columnwidth]{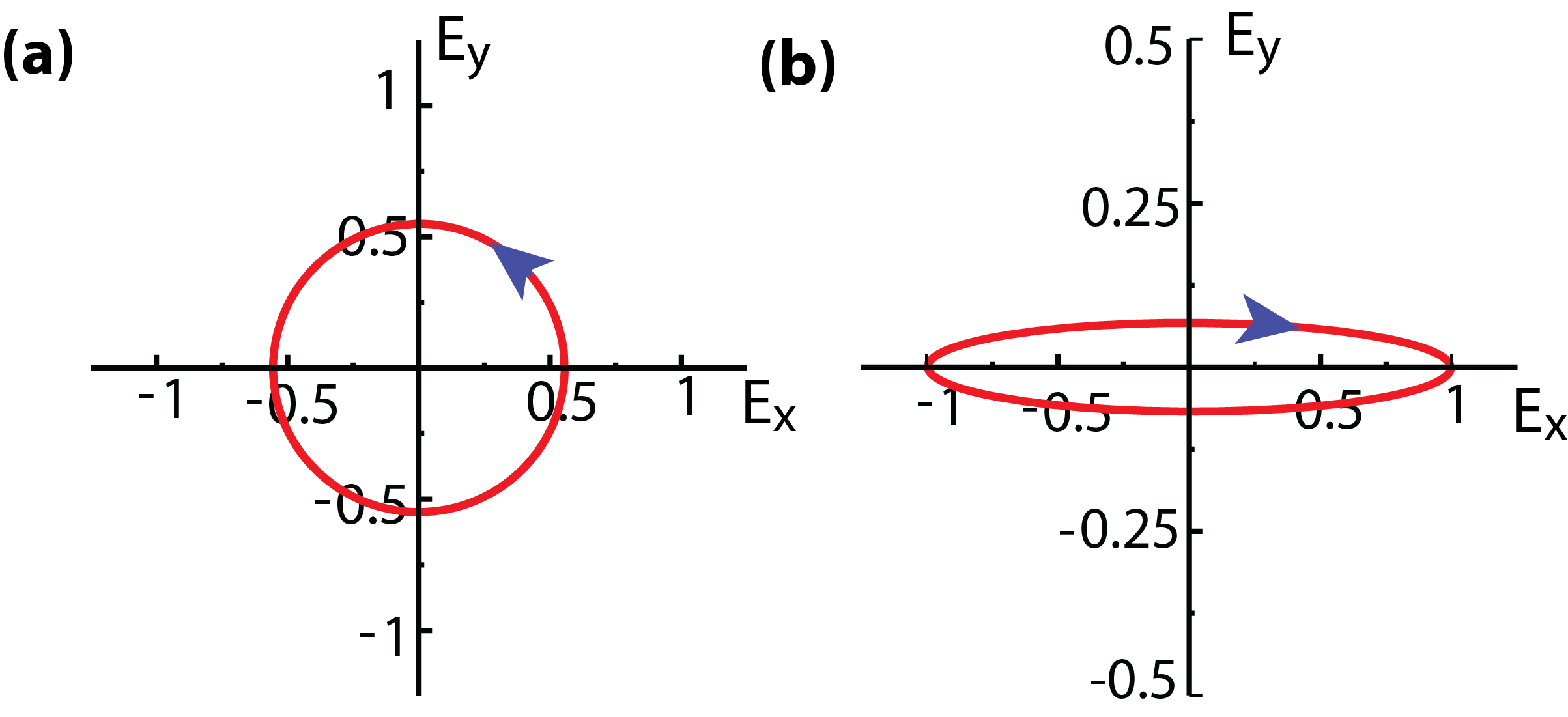}
\caption{Polarization ellipse of the far field radiation of topological spaser for its two regimes of continuous wave operation. (a) Radius of TMDC nanoflake is 16 nm and the gain is 49 $ps^{-1}$. Only $m=1$ plasmon mode is generated. The far field radiation is left circularly polarized. 
(b) Radius of TMDC nanoflake is 16 nm and the gain is 70 $ps^{-1}$. Two plasmon modes, $m=1$ and $m=-1$ are generated. The far field radiation is right elliptically polarized. The electric field is shown in arbitrary units. 
}\label{farfield}
\end{center}
\end{figure}

The components of the dipole radiated field are proportional to the total dipole moment, $E_x\propto \mathbf{d_{total,{x}}}$, $E_y\propto \mathbf{d_{total,{y}}}$. The corresponding polarization ellipse is shown in Fig.\ \ref{farfield} for two regimes of operation of the topological spaser: (a) only one $m=1$ plasmon mode is generated and (b) two modes, $m=1$ and $m=-1$, are cogenerated. For case (a), the far field radiation is left circularly polarized, which is the same polarization as the one of the pump light. This is consistent with the condition that only one plasmon mode is generated in this case. 

If two plasmon modes are generated [Fig. \ref{farfield}(b)], then the corresponding polarization ellipse describes the right elliptically polarized radiation. Note, that the pump light is left circularly polarized. The change of the handedness of polarization from left to right is due to the fact that the number of $m=-1$ plasmons is greater than the number of $m=1$ plasmons in the continuous wave regime of the spaser. 

Thus, for a given radius of TMDC nanoflake, by changing the gain, i.e., the intensity of the circularly polarized light, we can switch the handedness of the far field radiation from left to right and vica versa. 
For example, if $g_{th_1} <g< g_{th_2}$ then the far field radiation is left circularly polarized, but if $g_{th_2} <g$ then it is right elliptically polarized.

%The plot of the total dipole is shown in Fig. \ref{farfield}(a). The parametic graph is an elliptically polarized light due to the mixing of clockwise and anticlockwise components given by the the Eq.\eqref{19} and Eq.\eqref{20}. Fig. \ref{farfield}(b) represents parametric plot of the Magnetic field due to the dipole given by the Eq.\eqref{26} and Eq.\eqref{27}. This too is an elliptical rotated $\frac{\pi}{2}$ about the direction of the total dipole. Here, fitting was done for an ellipse which alligns perfectly with the actual datapoints as can be observed in the figure.

\section{Conclusion}
\label{Conclusion}

A topological nanospaser of type II consists of two main components: a metal nanospheroid and a TMDC, e.g., $\mathrm{MoS_2}$, monolayer flake of a circular shape. The nanospheroid functions as a plasmonic nanoresonator with two relevant plasmonic modes, which rotate in the opposite directions and are characterized by azimuthal quantum numbers $m = \pm 1$. 
The $\mathrm{MoS_2}$ monolayer is a gain medium with nontrivial topology. It is placed atop of a nanospheroid and has two  chiral $valleys$, $K$ and $K^\prime$. The system is pumped by a circular polarized light, which populates the conduction band states of only one valley, say the $K$ valley. 

Such topological spaser has been theoretically proposed in Ref. \cite{topological_nanospaser_rupesh}. In the present paper we show that it has very rich dynamics, which strongly depends on the radius of the gain medium (TMDC nanoflake).  If the radius of TMDC is small, then the $K$ and $K^\prime $ valleys are mainly coupled to the co-rotating plasmonic modes, e.g., the $K$ valley is coupled to the $m=1$ mode. In this case the nanospaser has one threshold, $g_{th}$, so that if the gain is larger than $g_{th}$ then the co-rotating plamon mode is generated. For larger radius of nanoflake, the valleys of TMDC become also strongly coupled to the counter-rotating modes, and the nanospaser has two thresholds, $g_{th,1}$ and $g_{th,2}$, so that if $g_{th,2}>g>g_{th,1}$ then only the co-rotating mode is generated, while if $g>g_{th,2}$ then both co-rotating and counter-rotating modes are generated. For even larger radius of TMDC, the two thresholds merge into one and the nanospaser has only one regime when two modes, $m=1$ and $m=-1$, are cogenerated.  In this case the number of counter-rotating plasmons is larger than the number of co-rotating ones. Because of that property the far-field radiation of nanospaser shows interesting behavior. Namely, by changing the gain strength, one can change the handedness of the far-field radiation from left to right and vice versa.

%We selectively pump $\mathbf{K}$ $valley$ and study the coupling between the plasmonic field and the transition dipole in the $valleys$. For a gain flake size of the footprint of spheroid, only a mode which matches the chirality of the valley i.e for our case $m=1$ is generated\cite{topological_nanospaser_rupesh}. However, if the fake size is sufficiently large the plasmonic field outside the the footprint of the disk polarizes the unpumped valley, i.e.  $\mathbf{K^\prime}$ $valley$ in our case. This brings about the generation of second mode($m = -1$) along with the expected one$ m = 1$. Apart from the near-field spasing, this nanospaser is an excellent source of far-field radiation which emits elliptically polarized light. 

All these unique properties of topological nanospaser make it an extremely viable option for several nanoscopic applications. Main areas are near-field spectroscopy and sensing where a plasmon frequency of a nanospaser can be tuned to work at the required condition. However, the topological nanospaser can also be used in optical interconnects and probing. Another key area is the biomedical one where similar systems have been previously adopted \cite{ Galanzha_Nat_Comm_Spaser_biological_probe_2017,biolaser} for the diagnosis and therupatics of cancer. With added topoligical chiral benefits, this nanolaser can be more effective in such detection. In addition to all these, the topological nanospaser has also a potential as an excellent far-field raditation source. 

\section{Acknowledgment}

Major funding was provided by Grant No. DE-SC0007043  from the Materials Sciences and Engineering Division of the Office of the Basic Energy Sciences, Office of Science, U.S. Department of Energy. Numerical simulations have been performed using support by
Grant No. DE-FG02-01ER15213 from the Chemical Sciences, Biosciences and Geosciences Division, Office of Basic Energy Sciences, Office of Science, US Department of Energy.

\section{Supporting Information}
\subsection{Metallic Oblate Spheroid: Geometry and Modes }
\label{Modes}

We consider an oblate spheroid, which in the Cartesian coordinate system is described by the following equation 
\begin{equation} 
\dfrac{x^2+y^2}{a^2}+\dfrac{z^2}{c^2}=1~,
\end{equation}
where  $a$ and $c$ are semi-axes, $\varepsilon=\sqrt{1-\frac{c^2}{a^2}}$ is the 
eccentricity of the spheroid. 

It is convenient to introduce the spheroidal coordinates, $\xi$, $\eta$ and $\varphi$, which are related to the Cartesian coordinates, $x$, $y$ and $z$ through the following expressio\cite{Willatzen_Voon_2011_Book_Boundary_Problems}:
\begin{align}
x&=f\sqrt{\xi^2+1}\sqrt{1-\eta^2}~ \cos(\varphi)\label{1},\\
y&=f \sqrt{\xi^2+1}\sqrt{1-\eta^2}~\sin(\varphi)\label{2},\\
z&=f \xi \eta\label{3},
\end{align}
where  $ 0 \leq \xi < \infty$, ~ $ -1 \leq \eta \leq 1$, ~ $ 0 \leq \varphi < 2 \pi$ and $f=\varepsilon a$.

Then the surface plasmon eigenmodes of the metal spheroid are described by the quasistatic equation \cite{Stockman:2001_PRL_Localization}
\begin{align}
\nabla\left[\theta(\mathbf{r})\nabla \phi_{m} \right]= s_{\mathrm{sp}} \nabla^2 \phi_{m}, 
\end{align}
where $s_{\mathrm{sp}}$ is the eigenvalue of the corresponding mode  $\phi_{m} $. 
Here $\theta(\mathbf{r})$ is the characteristic function that is 1 inside the metal and 0 elsewhere. 
For oblate spheroid, the eigenmodes are characterized by multipole quantum number $l$ and magnetic quantum number $m$. For the relevant modes of topological nanospaser, the  multipole quantum number is 1, $l=1$. Then the corresponding eigenmodes are described by the following expressions
\begin{align}
\phi_m =C_\mathrm{N}  P_{1}^{m}(\eta) e^{im\phi}
\begin{cases}
 \frac{P_1^{m}(i\xi)}{P_1^{m}(i\xi_{0})},  &  0<\xi<\xi_{0},\\
 \frac{Q_1^{m}(i\xi)}{Q_1^{m}(i\xi_{0})}, &  \xi_{0}<\xi,
\end{cases}\label{boundary}
\end{align}
where $ P_l^{m}(x)$ and $ Q_l^{m}(x)$ are the Legendre functions of the first and second kind, respectively, and 
$\xi_0 = \frac{\sqrt{1-\varepsilon^2}}{\varepsilon}$.
 The constant $C_\mathrm{N}$  is determined by normalization condition,
\begin{align}
\int_{\mathrm{All~Space}}  |\nabla \phi(\mathbf{r})_m|^2 d^3 \mathbf{r}=1.
\end{align}

Due to axial symmetry of the nanospheroid, the corresponding eigenvalues, $s_\mathrm{sp}$, do not depend on $m$. They can be also found from the following expression\cite{Bergman_Stockman:2003_PRL_spaser, Stockman_JOPT_2010_Spaser_Nanoamplifier}
\begin{align}
s_\mathrm{sp} = \frac{\displaystyle{ \int_{\mathrm{All~Space}}  \theta(\mathbf{r}) |\nabla \phi_m(\mathbf{r})|^2 d^3 \mathbf{r}}}{\displaystyle{ \int_{\mathrm{All~Space}}  |\nabla \phi_m(\mathbf{r})|^2 d^3 \mathbf{r}}}.
\label{s_sp}
\end{align}
Using explicit expression (\ref{boundary}) for $\phi_m$, we derive the final equation for the eigenvalue
\begin{align}
	s_\mathrm{sp} =\left. \frac	{\frac{d P_1^m(x)}{dx}} {\frac{d P_1^m(x)}{dx}  - \frac{P_1^m(x)}{Q_1^m(x)} \frac{d Q_1^m(x)}{dx}	}	\right|_{x=i\xi_0}~.
	\label{s_m}
\end{align}
	To find the plasmon frequency, $\omega_{\mathrm{sp}}$, and the plasmon relaxation rate, $\gamma_{\mathrm{sp}}$, we use the following relations  \cite{Bergman_Stockman:2003_PRL_spaser, Stockman_JOPT_2010_Spaser_Nanoamplifier}:
\begin{align}
&s_{\mathrm{sp}} = \mathrm{Re}[s(\omega_{\mathrm{sp}})], \\
&r_{\mathrm{sp}} = \frac{\mathrm{Im}[s(\omega_{\mathrm{sp}})]}{s_{{\mathrm{sp}}}^{\prime}},~~~~
s_{{\mathrm{sp}}}^{\prime}\equiv \frac{d\mathrm{Re}[s(\omega)]}{d\omega}\Big|_{\omega=\omega_{\mathrm{sp}}},
\end{align}
where the Bergman spectral parameter is defined as 
\begin{align}
s(\omega) = \frac{\epsilon_{d}}{\epsilon_{d}-\epsilon_{m}(\omega)}~.
\end{align}
Here $\epsilon_d$ is the dielectric constant of surrounding medium, and $\epsilon_m(\omega)$ is the dielectric function of the metal (silver). In our computations, for silver, we use the dielectric function from Ref.\ \cite{Johnson:1972_Silver}.

\subsection{TMDC Parameters}
\label{TMDC}

\begin{figure}
	\begin{center}
		\includegraphics[width=.95\columnwidth]{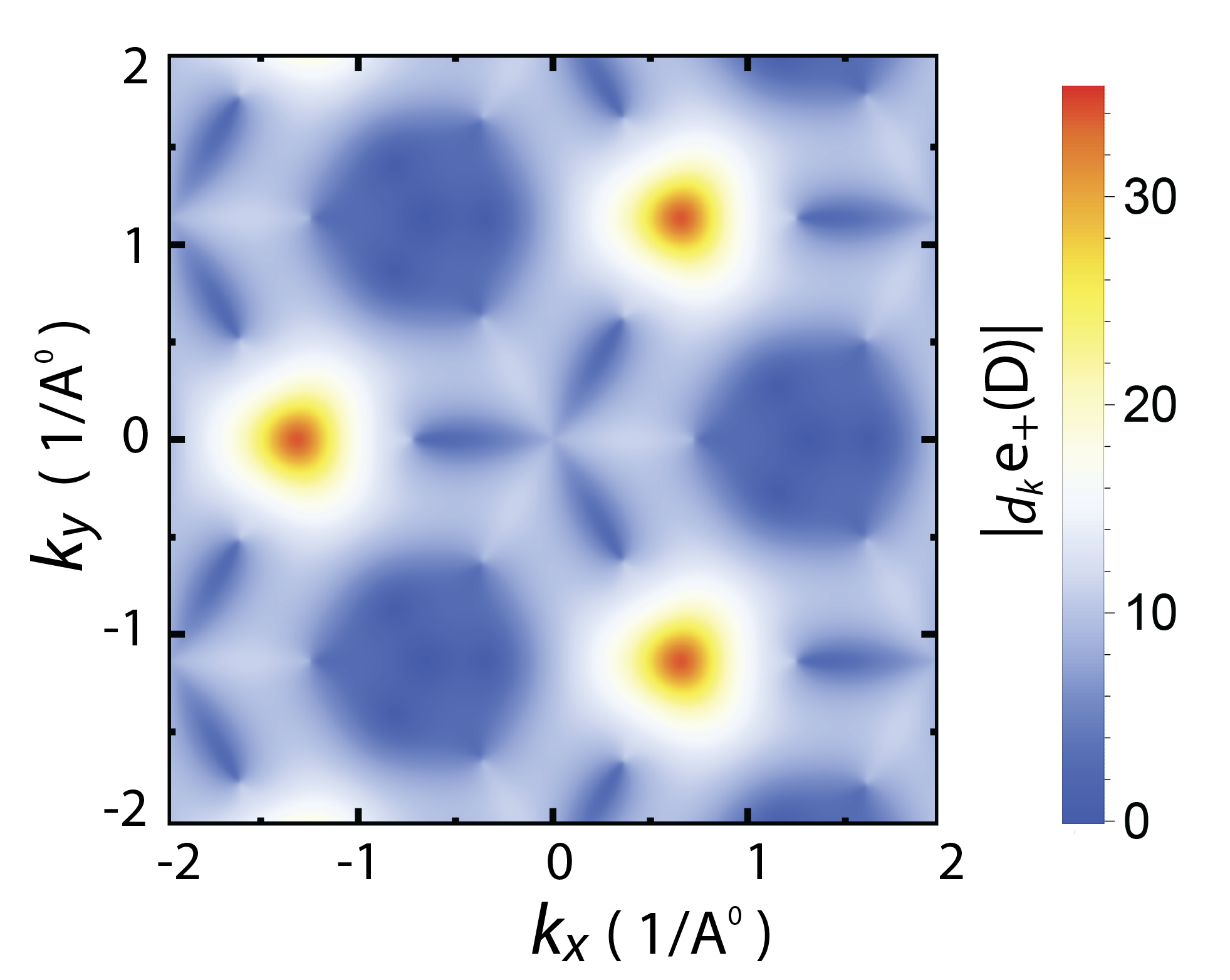}
		\caption{Absolute value of the left-rotating chiral dipole component, $\mathbf d_-=\mathbf e_+\mathbf d$, in MoS$_2$}
		\label{bist4.eps}
	\end{center}
\end{figure}
%%

%The eigenfrequency, $\omega_\mathrm{sp}$, depends on the aspect ratio, $c/a$. We set $a=12$ nm and vary the value of $c$ to have the surface plasmon frequency match the band gap, $ \omega_\mathrm{sp} = \Delta_\mathrm{g}$. 
%Table \ref{Table} shows the calculated  values of $c$ yielding this matching for the specific TMDCs.

In our calculations the TMDC ($\mathrm{MoS_2}$) monolayer is characterized by its bandgap and the dipole matrix elements between the conduction and valence bands at the $K$ and $K^\prime $ points. To find these parameters we have used a three-band tight binding model \cite{Liu_et_al_PRB_2014_Three_Band_Model}. The calculated values of the transition dipole matrix elements and the bandgap are given in Table \ref{Table}. The dipole matrix elements at the $K$ and $K^\prime$ points are purely chiral. They are proportional to $\mathbf e_\pm=2^{-1/2}\left(\mathbf e_x\pm i\mathbf e_y\right)$, where $\mathbf e_x$ and $\mathbf e_y$ are the Cartesian unit vectors. The plot of the absolute value of the chiral dipole, $\left|\mathbf d_\pm\right|$, where $\mathbf d_\pm=
\mathbf e_\pm^\ast \mathbf d$,  is shown in Fig. \ref{bist4.eps}.  

\begin{table}
\centering
	{\small
\begin{tabular}{ccccc}
%\begin{tabular}{ | c|>{\centering}p{2.2cm}| >{\centering}p{1.8cm}  |>{\centering}p{1.8cm} | c| l |}
\toprule
\multirow{2}{*} {TMDC} & 
		{Semi-principal axis } &
		 \multicolumn{2}{c|}{Dipole elements (D)} &
		 {Band gap} \\
  &  $c$ (nm) &  $\mathbf d_\mathbf K$  &  $\mathbf d_{\mathbf K^{\prime}}$ &(eV) \\ \hline
\midrule
$\mathrm{MoS}_2$   & $1.20$     &  $17.68 \mathbf{e}_{+}$  &  $17.68\mathbf{e}_{-}$                    &  1.66 \\ \hline
		$\mathrm{MoSe}_2$  & $1.45$    &  $19.23 \mathbf{e}_{+}$  &  $19.23\mathbf{e}_{-}$                     &1.79 \\ \hline
		$\mathrm{WSe}_2$   &  $0.85$   &  $18.38 \mathbf{e}_{+}$  &  $18.38\mathbf{e}_{-}$                      &1.43 \\ \hline
		$\mathrm{MoTe}_2$  &  $1$       &  $20.08 \mathbf{e}_{+}$  &  $20.08\mathbf{e}_{-}$                      & 1.53 \\ 
\bottomrule
\end{tabular}
}
\caption{Parameters employed in the calculations: Semi-principal axes of the spheroids, and the dipole matrix elements and band gaps of the TMDCs.}
\label{Table}
\end{table}

% By default, an article has some vary large margins to fit the smaller page format.  This allows us to use more standard margins.

\setlength{\parskip}{1em}
% This gives us a full line break when we write a new paragraph
% Once we have all of our packages and setting announced, we need to begin our document.  You will notice that at the end of the writing there is an end document statements.  Many options use this begin and end syntax.
\subsection{Far-field radiation}

%The total Hamiltonian is :
%\begin{align}
%H=H_{SP}+ H_{gain}+ H_{int}\\
%\mathrm{where},~
%H_{SP}=\hbar\omega_\mathrm{sp}\sum_{m=\pm1}  \hat{a}_m^{\dagger}\hat{a}_m^{},
%\end{align}

The total dipole moment of the spaser can be expressed in the following form 
\begin{align}
\mathbf{d_{total}}= \mathbf{d_{metal}}+\mathbf{d_{tmdc}}, \label{eq22}
\end{align}
where $\mathbf{d_{metal}}$ is the dipole moment of the metal nanospheroid and 
$\mathbf{d_{tmdc}}$ is the dipole moment of the TMDC nanoflake.

\subsubsection{Dipole moment of the metal nanospheroid}

The electric field inside the metal, which is produced by generated plasmon modes, both $m=1$ and $m=-1$, is uniform and is given by the following expression 
\begin{equation}
%\mathbf{F}_m(\mathbf{r})= - A_\mathrm{sp} \nabla \phi_m(\mathbf{r})(\hat{a}_m +\hat{ a}_m^{\dagger}),
\mathbf{F}_m(\mathbf{r},t)=-\sum_{m=1,{-1}}  A_\mathrm{sp} (\nabla \phi_m\hat{a}_m e^{-i\omega t} +\nabla \phi_m^{*}\hat{ a}_m^{*} e^{i\omega t}), \label{eq4}
\end{equation}
where 
\begin{equation}
A_ \mathrm{sp} =\sqrt{\frac{4\pi 
\hbar s({\omega)}}{\epsilon_d s^\prime(\omega)}} 
\end{equation}
and 
\begin{equation}
\mathrm{s(\omega)}=\frac{\epsilon_{d}}{\ 
\epsilon_{d}-\epsilon_{m(\omega)}}, 
\end{equation}
Then the dipole moment of the metal nanospheroid can be found from the following expression 
\begin{equation}
\mathbf{d_{metal}}=\int_V\mu \mathbf{F_m(\mathbf{r},t)}\,dv \label{eq5}
\end{equation}
where 
\begin{equation}
\mu  =\frac{\mathrm{Re}[\epsilon_{metal}-\epsilon_{d}]}{4\pi}.
%\mathbf{d_{metal}}=\mathbf{F}_m(\mathbf{r})(\frac{\mathrm{Re}[\epsilon_{metal}-\epsilon_{d}]}{4\pi}) \mathrm{V} \label{eq22}
\end{equation}
Taking into account that the electric field inside the metal is a constant, 
$E_0 = \lvert  \nabla \phi_m\rvert$, we derive the following expressions for the 
dipole moment of the metal nanospheroid
%$s_{m}=\bigintss  \lvert  \nabla \phi_m\rvert^{2} ~ d\mathrm{V}= \lvert  \nabla \phi_m\rvert^{2} \mathrm{V} =\mathrm{ E_{0}^{2}}\mathrm{V}$ and using eq (\ref{eq21}),  eq (\ref{eq5}) can be written as,
%\begin{align}
%%\mathbf{d_{metal}}^{2}=\frac{1}{4\pi}\frac{\hbar}{\frac{{\partial  s(\omega)}}{{\partial  \omega}}}\mathrm{Re}[\epsilon_{m}-\epsilon_{d}]^{2}\mathrm{V_{m}}\mathrm{N_{m}}
%\mathbf{d_{metal}}=\sqrt{\frac{\mathrm{V}}{4\pi}\frac{\hbar}{\frac{{\partial  s(\omega)}}{{\partial  \omega}}}}&\mathrm{Re}[\epsilon_{metal}-\epsilon_{d}]\bigg((\hat{a}_{1} e^{-i\omega t} +\hat{ a}_1^{*} e^{i\omega t})\\ \nonumber&+i(\hat{a}_{-1} e^{-i\omega t} -\hat{ a}_{-1}^{*} e^{i\omega t})\bigg) 
%\end{align}
%\mathbf{d_{metal}}=\sqrt{\frac{\mathrm{V}}{4\pi}\frac{\hbar}{\frac{{\partial  s(\omega)}}{{\partial  \omega}}}}&\mathrm{Re}[\epsilon_{metal}-\epsilon_{d}]\bigg((\hat{a}_{1} e^{-i\omega t} +\hat{ a}_1^{*} e^{i\omega t})\\ \nonumber&+i(\hat{a}_{-1} e^{-i\omega t} -\hat{ a}_{-1}^{*} e^{i\omega t})\bigg)
%\end{align}
\begin{align}
\mathbf{d_{metal,{x}}}=-\mu \mathrm{A}_{\mathrm{sp}}\mathrm{ E_{0} }\mathrm {V}&\bigg((\hat{a}_{1} e^{-i\omega t} +\hat{ a}_1^{*} e^{i\omega t})+\nonumber\\
&~~~~~~(\hat{a}_{-1} e^{-i\omega t} +\hat{ a}_{-1}^{*} e^{i\omega t})\bigg)\label{eq7}\\
\mathbf{d_{metal,{y}}}=-\mu \mathrm{A}_{\mathrm{sp}}\mathrm{ E_{0} }\mathrm {V}&\bigg(i(\hat{a}_{1} e^{-i\omega t} -\hat{ a}_1^{*} e^{i\omega t})-\nonumber\\
&~~~~~~i(\hat{a}_{-1} e^{-i\omega t} -\hat{ a}_{-1}^{*} e^{i\omega t})\bigg)\label{eq8} 
\end{align}

\subsubsection{Dipole moment of TMDC monolayer}

The density matrix of TMDC nanoflake has the following structure
\begin{align}
\hat{\rho}_{\mathbfcal{K}}(\mathbf{r},t) & =\left(
\begin{array}{cc}
\rho_{\mathbfcal{K}}^\mathrm{(c)}(\mathbf{r},t) & \rho_{\mathbfcal{K}}(\mathbf{r},t) e^{i\omega t}\\
\rho_{\mathbfcal{K}}^{*}(\mathbf{r},t)  e^{-i\omega t} & \rho_{\mathbfcal{K}}^\mathrm{(v)}(\mathbf{r},t)
\end{array}
\right).
\end{align}
where  ${\mathbfcal{K}}$ is the valley index, $K$ or $K^\prime$. The off-diagonal elements, i.e., coherences, determine the dipole moment of TMDC system
\begin{equation}
\mathbf{d_{tmdc}} =\sum_{\mathrm{S}} 
\sum_{\mathbfcal{K}=\mathbf{K},{\mathbf{K}^\prime}}
(\rho_{\mathbfcal{K}}(\mathbf{r}) \mathbf{d_{\mathbfcal{K}}}e^{i\omega t}+\rho_{\mathbfcal{K}}^{*}(\mathbf{r})
\mathbf{d_{\mathbfcal{K}}^{*}}e^{-i\omega t}) + h.c.,
\label{eqtmdc}
\end{equation}
where $\sum_{\mathrm{S}}$ is the sum (integral) over all points $\mathbf{r}$ 
of TMDC nanoflake. 

The coherences satisfy the following stationary equation (see Eq. (12) of the main text)
\begin{align}
[-i(\omega-\omega_{21})-\Gamma_{12}] {\rho}_{\mathbfcal{K}}(\mathbf{r})+
i n_{\mathbfcal{K}}(\mathbf{r})\sum_{m=1,{-1}} \tilde{\Omega}_{m,\mathbfcal{K}}^{*}(\mathbf{r}) a_{m}^{*} = 0,
\label{14}
\end{align} 
where  $\Gamma_{12}$ is the polarization relaxation rate, $n_{\mathbfcal{K}}$
is the population inversion defined as
\begin{equation}
n_{\mathbfcal{K}} \equiv \rho_{\mathbfcal{K}}^\mathrm{(c)} - \rho_{\mathbfcal{K}}^\mathrm{(v)}~,
\end{equation}
and
\begin{align}
\tilde{\Omega}_{m,\mathbfcal{K}}(\mathbf{r})&= -\frac{1}{\hbar} A_\mathrm{sp} \nabla \phi_m(\mathbf{r})\mathbf{d}_\mathbfcal{K}~.
\label{Rabi}
\end{align}
From Eq. (\ref{14}) we can find the stationary coherences of TMDC monolayer
\begin{align}\rho_{\mathbfcal{K}}(\mathbf{r})= -\frac{i n_{\mathbfcal{K}}(\mathbf{r})\sum_{m=1,{-1}} \tilde{\Omega}_{m,\mathbfcal{K}}^{*} (\mathbf{r}) a_{m}^{*}}{-(\omega-\omega_{21})+i\Gamma_{12}}. \label{rho1}
\end{align}
We substitute Eq. (\ref{rho1}) into Eq. (\ref{eqtmdc}) and obtain the following expression for the dipole moment of TMDC
\begin{equation}
\mathbf{d_{tmdc}}=\mathrm{f}_{\mathbf{K}}\mathbf{d_{K}}e^{i\omega t}+
\mathrm{f}_{\mathbf{K^\prime}}\mathbf{d_{K^\prime}}e^{i\omega t}+ h.c.,
\end{equation}
where the following notations were introduced 
\begin{align}
\mathrm{f}_{\mathbf{K}}=- {\nu}&\sum_{\mathrm{S}}\frac{in_{\mathbf{K}}(\mathbf{r})\sum_{m=1,{-1}} \tilde{\Omega}_{m,\mathbf{K}}^{*}a_{m}^{*}}{-(\omega-\Delta_\mathrm g)+i\Gamma_{12}}\\
\mathrm{f}_{\mathbf{K^{\prime}}}=- {\nu}&\sum_{\mathrm{S}}\frac{in_{\mathbf{K^{\prime}}}(\mathbf{r})\sum_{m=1,{-1}} \tilde{\Omega}_{m,\mathbf{K^{\prime}}}^{*}a_{m}^{*}}{-(\omega-\Delta_\mathrm g)+i\Gamma_{12}}.
\end{align}
Taking into account that $\mathbf{d_{K}}=\mathrm{d_{0}}(1,i)$ and $\mathbf{d_{K^\prime}}=\mathrm{d_{0}}(1,-i)$ we obtain the $x$ and $y$ components of the dipole moment 
\begin{align}
\mathbf{d_{tmdc,{x}}}&=\mathrm{f}_{\mathbf{K}}\mathrm{d_{0}}e^{i\omega t}+
\mathrm{f}_{\mathbf{K^\prime}}\mathrm{d_{0}}e^{i\omega t}+ h.c. \label{eq21}\\
\mathbf{d_{tmdc,{y}}}&=i\mathrm{f}_{\mathbf{K}}\mathrm{d_{0}}e^{i\omega t}-i
\mathrm{f}_{\mathbf{K^\prime}}\mathrm{d_{0}}e^{i\omega t} + h.c. \label{eq211}
\end{align}

\subsubsection{Far field dipole radiation}

The total dipole moment of the spaser system is the sum of the dipole moment of the metal nanospheroid and TMDC nanoflake. Its $x$ and $y$ components can be expressed as 
\begin{align}
\mathbf{d_{total,{x}}}=-\mu \mathrm{A}_{\mathrm{sp}}\mathrm{ E_{0} }\mathrm {V}&\left(  \hat{a}_{1} e^{-i\omega t} +\hat{a}_{-1} e^{-i\omega t} + h.c.\right)\nonumber\\
&+\left( \mathrm{f}_{\mathbf{K}}\mathrm{d_{0}}e^{i\omega t}+
\mathrm{f}_{\mathbf{K^\prime}}\mathrm{d_{0}}e^{i\omega t}+ h.c. \right) 
 \\
\mathbf{d_{total,{y}}}=-\mu \mathrm{A}_{\mathrm{sp}}\mathrm{ E_{0} }\mathrm {V}&\left((i\hat{a}_{1} e^{-i\omega t}-i\hat{a}_{-1} e^{-i\omega t} +h.c. \right)
\nonumber\\
&+\left( i\mathrm{f}_{\mathbf{K}}\mathrm{d_{0}}e^{i\omega t}-i
\mathrm{f}_{\mathbf{K^\prime}}\mathrm{d_{0}}e^{i\omega t} + h.c. \right) 
\label{eq211}
\end{align}
These expressions have the following structure
\begin{align}
\mathbf{d_{total,{x}}}&=2 \mathrm{Re} \left[ B_x e^{i\omega t}  \right] ,\\
\mathbf{d_{total,{y}}}&=2 \mathrm{Re} \left[ B_y e^{i\omega t}  \right],
\end{align}
where,
\begin{align}
B_x&=-\mu \mathrm{A}_{\mathrm{sp}}\mathrm{ E_{0} }\mathrm {V}(\hat{ a}_1^{*}+\hat{ a}_{-1}^{*})+\mathrm{f}_{\mathbf{K}}\mathrm{d_{0}}+\mathrm{f}_{\mathbf{K^\prime}}\mathrm{d_{0}},\\
B_y&=i\mu \mathrm{A}_{\mathrm{sp}}\mathrm{ E_{0} }\mathrm {V}(\hat{ a}_1^{*}-\hat{ a}_{-1}^{*})+i\mathrm{f}_{\mathbf{K}}\mathrm{d_{0}}-i\mathrm{f}_{\mathbf{K^\prime}}\mathrm{d_{0}}.
\end{align}

%Thus, 
%\begin{align}
%\langle \mathrm{\tilde {d_x}d_x+\tilde {d_y}d_y}\rangle=(\mathrm{\tilde B_1} e^{i\omega t}&+{\mathrm{B_1} }e^{-i\omega t})(\mathrm{B_1} e^{i\omega t}+%\tilde{\mathrm{B_1} }e^{-i\omega t})\nonumber\\
%&+(\mathrm{\tilde C_1} e^{i\omega t}+{\mathrm{C_1} }e^{-i\omega t})(\mathrm{C_1} %\end{align}
%Since, $\langle\mathbf{e^{\pm i\omega t}\rangle}=0$, equation \ref{eq29} can be written as:

%\begin{align}
%\langle \mathrm{\tilde {d_x}d_x+\tilde {d_y}d_y}\rangle=2\lvert {\mathrm{B_1} }%\rvert^2+2\lvert {\mathrm{C_1} }\rvert^2\label{eq30}
%\end{align}
%\paragraph{Radiation Rate:}

The total dipole moment of the system determines the far-field radiation of the spaser. The polarization of radiation is characterized by the $x$ and $y$ components of the far electric field, which are proportional to the corresponding components of the dipole moment, i.e., $\mathbf{d_{total,{x}}}$ and $\mathbf{d_{total,{y}}}$, while the total radiation power is given by the following expression 
\begin{align}
I&=\frac{4}{3} \left(\frac{\omega}{c_0}\right)^3\frac{ \left(\epsilon_\mathrm d\right)^{1/2}}{\hbar}\langle |\mathbf{d_{total}}|^2 \rangle\nonumber\\
&=\frac{8}{3} \left(\frac{\omega}{c_0}\right)^3\frac{ \left(\epsilon_\mathrm d\right)^{1/2}}{\hbar}(\lvert {B_x }\rvert^2+\lvert { B_y }\rvert^2)
\end{align}
%\paragraph{Expression for $ \mathrm{A}_{\mathbf{sp}}\mathbf{ E_{0} }$:}
%\begin{align}
%\mathrm{A}_{\mathrm{sp}}\mathrm{ E_{0} } =\sqrt{\frac{4\pi 
%\hbar s_{n}}{\epsilon_d {\frac{{\partial  s(\omega)}}{{\partial  \omega}}}}}\times \sqrt{\frac{s_{n}}{\mathrm{V}}} , \label{eq32}
%\end{align}
where $\langle  \ldots \rangle$ means the time average. 
%
%\bibliography{references}
%%\bibliographystyle{plainnat}
%%\bibliographystyle{ieeetr}
%\bibliographystyle{plain}

\end{document}